\documentclass[nofootinbib,amsmath,amssymb,twocolumn,aps,pra,10pt,floatfix,superscriptaddress,a4paper]{revtex4-1}

\usepackage{graphicx} 

\usepackage{float} 
\usepackage{wrapfig} 
\usepackage{lipsum} 

\usepackage[utf8]{inputenc}
\usepackage{amsmath}
\usepackage{amsfonts}
\usepackage{amssymb}
\usepackage{fancyhdr}
\usepackage{amsthm}

\usepackage{enumitem}
\usepackage{amsmath}
\usepackage{braket}
\usepackage{graphicx}
\usepackage{pgfplots}

\usepackage{placeins}

\graphicspath{ {pics/} }

\graphicspath{{Pictures/}} 

\DeclareMathOperator{\Tr}{Tr}

\graphicspath{ {pics/} }

\newcommand{\crop}[2]{{\hat{#1}}_{#2}^{\dagger}}

\newcommand{\hilbert}{\mathcal{H}}
\newcommand{\repU}{\mathcal{U}}

\newcommand{\textin}{\text{in}}

\newcommand*{\QED}{\hfill\ensuremath{\square}}%

\theoremstyle{definition}

\newtheorem{conjecture}{Conjecture}[section]

\theoremstyle{plain}
\newtheorem{theorem}{Theorem}[section]
\newtheorem{lemma}[theorem]{Lemma}
\newtheorem{proposition}[theorem]{Proposition}
\newtheorem{corollary}[theorem]{Corollary}

\theoremstyle{remark}

\newcommand{\avgbracket}[1]{\langle #1 \rangle}

\begin{document}
\title{Generating entanglement with linear optics}
\author{Stasja \surname{Stanisic}}
\affiliation{Quantum Engineering Technology Labs, H. H. Wills Physics Laboratory and Department of Electrical \& Electronic Engineering, University of Bristol, UK}
\affiliation{Quantum Engineering Centre for Doctoral Training, University of Bristol, UK}
\author{Noah \surname{Linden}}
\affiliation{School of Mathematics, University of Bristol, UK}
\author{Ashley \surname{Montanaro}}
\affiliation{School of Mathematics, University of Bristol, UK}
\author{Peter S. \surname{Turner}}
\affiliation{Quantum Engineering Technology Labs, H. H. Wills Physics Laboratory and Department of Electrical \& Electronic Engineering, University of Bristol, UK}

\date{\today}

\begin{abstract}
Entanglement is the basic building block of linear optical quantum computation, and as such understanding how to generate it in detail is of great importance for optical architectures.
We prove that Bell states cannot be generated using only 3 photons in the dual-rail encoding, and give strong numerical evidence for the optimality of the existing 4 photon schemes.
In a setup with a single photon in each input mode, we find a fundamental limit on the possible entanglement between a single mode Alice and arbitrary Bob.
We investigate and compare other setups aimed at characterizing entanglement in settings more general than dual-rail encoding.
The results draw attention to the trade-off between the entanglement a state has and the probability of postselecting that state, which can give surprising constant bounds on entanglement even with increasing numbers of photons.
\end{abstract}


\maketitle

\section{Introduction}
\label{sec:introduction}

Research into quantum technologies has gained significant momentum in the last several years, with applications ranging across metrology, communications, security, simulation and computation \cite{OBrien2010, Saffman2010, Xiang2013, Kok2007b}.
One of the important resources lying behind many of these advances is quantum entanglement \cite{Jozsa2003, Horodecki2009}.
Long before it was a potential technological resource, entanglement was studied as one of the phenomena lying at the foundations of quantum mechanics \cite{Bell1964,Clauser1969,Popescu1994}.
That there exist non-classical correlations between physical systems is now well established, while how best to generate, verify and quantify such entangled states in practice is an ongoing field of activity.
What is practical in any given situation depends on the physical platform under consideration; here we will be interested in the generation of entanglement using linear optics and postselection.\par

In linear optics we study collections of optical modes, modelled as harmonic oscillators whose excitations correspond to photons.
Interactions are restricted to Hamiltonians that leave the total number of photons fixed, giving rise to unitary transformations on modes (interferometers), as well as possible measurement and postselection of quantum states (heralding).
This realization introduces an interesting set of constraints on the entanglement problem.
Most work to date focuses on either single- or dual-rail encoding of photons into two-dimensional qubits, and then applying the usual approaches to quantum computation such as the circuit model or measurement-based schemes.
Gates are carried out via ancilla modes and photon detection measurements \cite{Kok2007b}.
The dual-rail encoding, where qubits are realized as single photons in pairs of spatial or polarization modes, is the commonly accepted standard for quantum computation with linear optics, and allows us to discuss entanglement in terms of standard concepts such as Bell and GHZ states \cite{Kok2007b, Carolan2015, Gimeno-Segovia2015}.
However, the requirement of postselection means generation of such states is nondeterministic, and the probability of success is often low; for example, the best known Bell state generation scheme has success probability of $1/4$ \cite{Joo2007} and if the postprocessing technique known as procrustean distillation is not allowed, then the probability drops to $0.1875$ \cite{Zhang2008}.
When we consider the number of Bell states needed to construct two-dimensional cluster states \cite{Gimeno-Segovia2015}, the requirements can be quite daunting, though promising proposals exist \cite{Rudolph2016}.
This helps to motivate the study of entanglement generation in linear optics more generally; in particular, it is natural to consider entanglement between two subsets of modes, foregoing encoding altogether.
While this is currently not the preferred way of generating entanglement, any bounds that can be found present fundamental limits on linear optical architectures, as well as for other quantum information processing tasks such as boson sampling \cite{Aaronson2010}.
\par

\begin{table*}[!htbp]
\centering
\begin{tabular}{|l|l|l|l|}
\hline
Bound (ebits) & Parameters & Input state & Section  \\ \hline
$ O(\log{n}) $   		&  $ M_A = 1, M_B = 1, M_H = 0    $ 				  & Bunched    &  \ref{subsec:bunched}              \\ \hline
$  2   $ 				&  $ M_A = 1, M_B \geq 1, M_H = 0 $					  & Unbunched  &  \ref{subsec:oppimConstant}              \\ \hline
$ \log{3} $     		&  $ M_A = M_B = 1, M_H \geq 1    $ 			  	  & Unbunched  & \ref{subsec:oppimMeasurement}               \\ \hline
$  \log{\left( 2 {M_A + \frac{n - 1}{2} \choose M_A } \right)}  $ 
                                &  $ M_A = M_B     $, $n$ odd    		  & Any        &  \ref{subsec:dimensionality}              \\ \hline
$  \log{\left( 2 \frac{n + M_A}{n} { M_A + \frac{n}{2} - 1 \choose M_A} \right)}$  
								&  $ M_A = M_B     $, $n$ even   	      & Any        &  \ref{subsec:dimensionality}              \\ \hline
$  n   $ 				&  $ M_A = M_B     $    		   					  & Fock state & \ref{subsec:linearity}               \\ \hline
\end{tabular}
\caption{
Entanglement bounds proven in this paper. The notation is as defined in Section \ref{sec:background} (see Figure \ref{fig:setup}).
}
\label{table:bounds}
\end{table*}

\par
In this paper we will consider two main themes regarding bipartite entanglement in linear optics; that where the parts are encoded qubits, and that where they are collections of modes.
Section \ref{sec:background} introduces the background and notation used throughout.
Section \ref{sec:qubit} examines qubit entanglement within the standard linear optical dual-rail encoding.
When we speak of dual-rail encoding, we mean qubit states that are post-selected such that there is exactly one photon in each pair of modes.
First we prove that one cannot generate a Bell state using only 3 photons, and then we give strong numerical evidence for the known 4 photon Bell state generator (with a success probability of $0.1875$) being optimal.
In Section \ref{sec:comparison} we compare qubit and mode entanglement, including an investigation of the expected average entanglement over uniformly (Haar) distributed interferometers.
In Section \ref{sec:modeEntanglement} we shift our focus to mode entanglement, considering bipartite systems made from two sets of optical modes, Alice and Bob, with a fixed total number of photons.
We see two types of behaviour. In the case of bunched photon input and single mode Alice, we find the entanglement can grow as $\log{n}$ where $n$ is the number of photons.
On the other hand, looking at the case of at most a single photon per input mode (as in, for example, boson sampling \cite{Aaronson2010}), a single mode Alice and no measurement, the entanglement is bounded above by $2$~ebits regardless of how many photons are present.
If we also restrict Bob to a single mode and furnish the remaining modes with number resolving detectors, the expected entanglement is bounded by $\log{3}$~ebits.
We then find provable universal bounds on the mode entanglement stemming from the dimensionality of the bipartite Fock states involved, and from the linearity of the optical transformation.
Finally, we conjecture a third bound due to unitarity which extends the previously mentioned constant bound in the case of Alice having a single mode to multi-mode Alice, and we provide numerical evidence for this conjecture.
The maximum mode entanglement is summarized in Table \ref{table:bounds}.

\section{Background}
\label{sec:background}

\begin{figure}[ht]
    \centering
	\includegraphics[width=.45\textwidth]{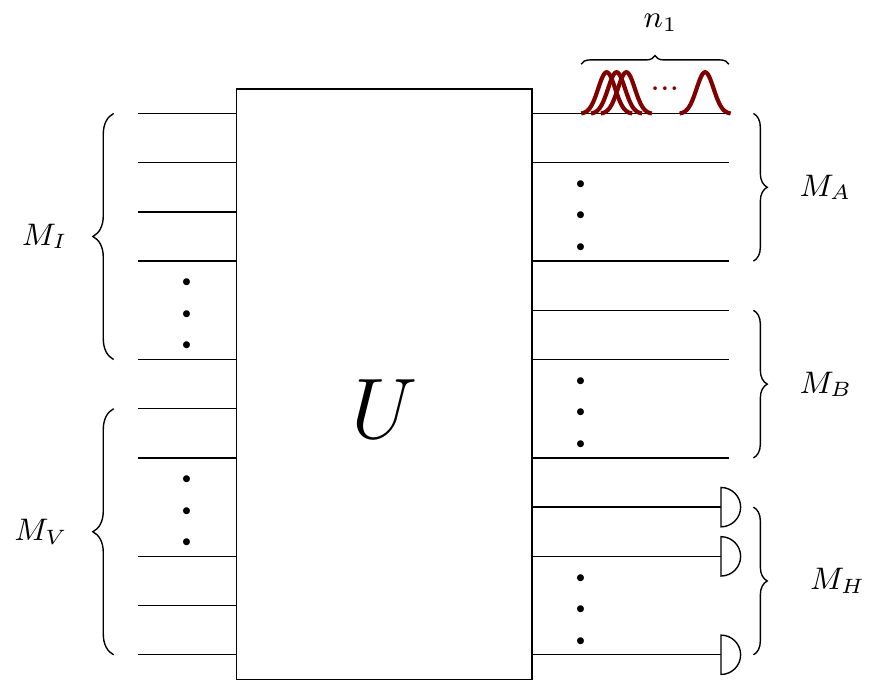}
	\caption{The generic setup used throughout this paper; see text for an explanation of the notation.
	}
	\label{fig:setup}
\end{figure}

Figure~\ref{fig:setup} introduces the generic linear optical setup and notation used throughout the paper.
The interferometer has $M$ input modes and $M$ output modes.
The mode transformation describing this (photon number preserving) interferometer is an $M \times M$ unitary matrix $U \in$ U($M$).
The top $M_I$ input modes contain $n$ input photons, while the bottom $M_V$ modes are ancilla vacua.
The representation of $U$ carried by the $n$ photon, $M$ mode Hilbert space in the number state (Fock) basis is denoted $\repU^{(n)}$.
The top $M_A$ output modes belong to one party, Alice, the middle $M_B$ modes belong to Bob, and the bottom $M_H$ modes -- Harold -- get measured using photon counting detectors.
Harold's detection pattern is labelled $\underline{h} = (n_{M_{S} + 1}, \cdots ,n_M)$ where $n_i$ gives the photon number of output mode $i$, and $M_S=M_A+M_B$ is the number of modes in the ``system'', i.e. modes that do not belong to Harold and are therefore unmeasured.
If $n_H=\sum_{k = M_{S+1}}^{M} n_k = ||\underline{h}||_1$ total photons have been detected, the number of photons left in the system is $n_S = n - n_H = n_A + n_B$.
The Hilbert space of subsystem $X$ (a subset of modes), given that it contains exactly $n_X$ photons, is denoted $\hilbert_X^{n_X}$.\par
	
Let the input to the interferometer be a Fock state
\begin{align}
\ket{\psi_{\textin}}
 &= \ket{n_1, n_2, \cdots, n_{M_I}, \underbrace{0, \cdots, 0}_{M_V}} \\
 &= \prod_{k = 1}^{M_I} \frac{(\crop{a}{k})^{n_k}}{\sqrt{n_k !}} \ket{\mathrm{vac}} ,
\end{align}
which transforms according to
\begin{align}
\repU^{(n)} \ket{\psi_{\textin}}
 &=  \prod_{k = 1}^{M_I} \frac{1}{\sqrt{n_k !}} \left( \repU^{(n)} \crop{a}{k} {\repU^{(n)}}^\dag \right)^{n_k} \repU^{(n)} \ket{\mathrm{vac}} \\
 &=  \prod_{k = 1}^{M_I} \frac{1}{\sqrt{n_k !}} \left( \sum_{j=1}^{M} \crop{a}{j} U_{jk} \right)^{n_k} \ket{\mathrm{vac}} , \label{eq:alg}
\end{align}
where $U_{jk}$ are the matrix elements of the mode transformation $U$, $\repU^{(n)}$ is the representation of $U$, and we have used the fact that the vacuum is invariant under all such transformations.
We will usually be interested in the case of single photon Fock inputs, where $n_k=1$ or vacuum for all input modes $k$, a situation we will refer to as unbunched.
If all the photons are found in one mode and the rest contain vacuum, we will refer to the state as completely bunched.
\par
When $M_H>0$ the ideal number resolving detectors will register a detection pattern $\underline{h} = (n_{M_{S} + 1}, \cdots ,n_M)$ of $n_H$ photons.
The output will be the post-measurement state consisting of $n_S=n-n_H$ photons remaining in the system modes $1, \cdots, M_S$, given by
\begin{align}
\ket{\psi_{S} (\underline{h}, U)}
 = \frac{\braket{ \underline{h} | \repU^{(n)} |\psi_{\textin}}}{\|  \braket{\underline{h}| \repU^{(n)} | \psi_{\textin}} \|} .
\label{eq:state}
\end{align}
Note that this is a pure state on the system $S=AB$, because $\ket{\underline{h}}$ only has support on subsystem $H$.
We will denote the unnormalized output by $\ket{\widetilde{\psi}_{S} (\underline{h}, U)} = \braket{ \underline{h} | \repU^{(n)} |\psi_{\textin}}$.
The Hilbert space of the system is
\begin{equation}
\hilbert_S^{n_{S}} = \bigoplus_{n_A = 0}^{n_S} \hilbert_A^{n_{A}} \otimes \hilbert_B^{n_{B}} ,
\label{eq:hilbertns}
\end{equation}
where $n_B = n_S - n_A$ is the number of photons in Bob's subsystem.
We are interested in entanglement with respect to this tensor product structure.
The dimension of the Hilbert space of $n$ photons in $M$ modes is $\binom{M + n - 1}{n}$, and so
\begin{align}
\dim \hilbert_{S}^{n_S} & = \sum_{n_A = 0}^{n_S} \binom{M_A + n_A - 1 }{ n_A} \binom{M_B + n_B - 1 }{n_B} \\ &= \binom{M_S + n_S - 1 }{ n_S}  \label{eq:dimAB}
\end{align}
as $M_S = M_A + M_B$ and $n_S = n_A + n_B$.
The totality of states available to Alice can be thought of as the Hilbert space $\bigoplus_{n_A=0}^{n_S} \hilbert_A^{n_A}$, and we may index its Fock basis as $\{ \ket{\underline{a}}_A \,:\, \underline{a}=(n_1, n_2, \cdots, n_{M_A}), ||\underline{a}||_1=n_A  \}$.
Similarly for Bob.
Expanding the output in this basis, we have
\begin{align}
\ket{\widetilde{\psi}_{S} (\underline{h}, U)} = \sum_{\underline{a},\underline{b}} \widetilde{C}_{\underline{a},\underline{b}}(\underline{h},U) \ket{\underline{a}}_A \otimes \ket{\underline{b}}_B . \label{eq:sd}
\end{align}
The coefficients $\widetilde{C}$ are related to permanents of the matrix $U$ \cite{Scheel2004, Aaronson2010}.
More specifically, consider an input Fock state $\ket{\psi} = \ket{n_1 \cdots n_M}$ and an output Fock state $\ket{\phi} = \ket{n'_1 \cdots n'_M}$ both with a total number of $n$ photons.
Construct a new matrix $U_{\psi \phi}$ from $U$ in two steps.
First, define the matrix $U_{\psi}$ consisting of $n_j$ copies of the $j$-th column of $U$ for all $j \in \{1, \cdots ,M\}$.
Next, construct the matrix $U_{\psi \phi}$ by using $n'_j$ copies of the $j$-th row of $U_{\psi}$ for all $j \in \{1, \cdots ,M\}$.
Then
\begin{equation}
\bra{\phi} \repU^{(n)} \ket{\psi} = {\frac{\text{perm}(U_{\psi \phi})}{\sqrt{n_1! \cdots n_M! n'_1! \cdots n'_M!}}}.
\end{equation}
In our notation, $\ket{\psi} = \ket{\psi_\mathrm{in}}$ and $\ket{\phi} =\ket{\underline{a}\underline{b}\underline{h}}$, we therefore have $\widetilde{C}_{\underline{a},\underline{b}}(\underline{h},U) = \bra{\underline{a}\underline{b}\underline{h}} \repU^{(n)} \ket{\psi_\mathrm{in}}$. 
The probability of detecting pattern $\underline{h}$ is $P(\underline{h}, U) = \sum_{\underline{a},\underline{b}} | \widetilde{C}_{\underline{a},\underline{b}}(\underline{h}, U)|^2$, and defining $C_{\underline{a},\underline{b}} = \widetilde{C}_{\underline{a},\underline{b}} / \sqrt{P(\underline{h}, U)}$, the normalized state can be written as $\ket{\psi_{S} (\underline{h}, U)} = \sum_{\underline{a},\underline{b}} C_{\underline{a},\underline{b}}(\underline{h}, U) \ket{\underline{a}}_A \ket{\underline{b}}_B$.
\par
For future convenience we define coefficients of the output states in particle notation, where the Fock state $\ket{n_1 \cdots n_M}$ is written as $\ket{\underbrace{1 \cdots 1}_{n_1}\cdots  \underbrace{M \cdots M}_{n_M}}$.
We denote relevant coefficients in particle notation by $\gamma$, which are related to the above mentioned permanent as
\begin{equation}
\gamma_{1 \cdots 1 \cdots M \cdots M}(\underline{h}, U) = \frac{\widetilde{C}_{\underline{a},\underline{b}}(\underline{h}, U) }{\sqrt{n_1 ! ... n_{M_S} !}} .
\end{equation}
These are the coefficients of the output states as expressed in terms of the creation operators assuming unbunched input to the interferometer, see Eq.(\ref{eq:bellcoef}).
\par
Equation~(\ref{eq:sd}) provides a Schmidt decomposition we can use to compute the entanglement.
However, the fact that the total number of photons in the system, $n_S$, is preserved implies that not all conceivable bipartite basis states $\ket{\underline{a}}_A \otimes \ket{\underline{b}}_B$ are available, so the system should not simply be viewed as the tensor product of two qudits i.e. Eq.(\ref{eq:dimAB}) is not simply the product of dim$\hilbert_A$ and dim$\hilbert_B$.
In particular, this means that states that are maximally entangled in the usual sense do not exist.
For example, Alice can have many states with $n_S$ photons, but there is only one possible Bob state to which they can be correlated, namely the vacuum (see Section \ref{subsec:dimensionality}).
\par
The entanglement measure that will be used is the von Neumann entropy; given a pure state $\ket{\psi_{S} (\underline{h}, U)}$, its density matrix is defined $\rho_{AB} (\underline{h}, U) =  \ket{\psi_{S} (\underline{h}, U)} \bra{\psi_{S} (\underline{h}, U)}$, and its reduced density matrices on subsystems are the marginals $\rho_{A} (\underline{h}, U) = \Tr_B [\rho_{AB} (\underline{h}, U)]$ and $\rho_{B} (\underline{h}, U)  = \Tr_A [\rho_{AB} (\underline{h}, U)]$.
The von Neumann entropy is then $S(\rho_A (\underline{h}, U) ) = - \Tr [{\rho_A (\underline{h}, U)  \cdot \log \rho_A (\underline{h}, U) }] = - \sum_{a} {\lambda_a \cdot \log \lambda_a} $ where $\{ \lambda_a \}_a$ are the non-zero eigenvalues of the reduced state.
Unless stated otherwise, logarithms will be assumed to be base $2$.
Finally, we will use ebits as the unit of bipartite entanglement where $1$ ebit corresponds to the von Neumann entropy of a Bell state.

\section{Qubit entanglement}
\label{sec:qubit}

In this section we will be considering the dual-rail encoding of two qubits.
This means that $M_A = M_B = 2$ and states are postselected so that subsystems $A$ and $B$ have exactly one photon each, $n_A=n_B=1$; all the other states are discarded.
(In general, the $k$-th qubit consists of the modes $2k-1$ and $2k$ via the mapping $\ket{10}_{2k-1,2k} \rightarrow \ket{0}_{k}$, $\ket{01}_{2k-1,2k} \rightarrow \ket{1}_{k}$.)
Despite the full Hilbert space of the system being of dimension $10$ (see Eq.~(\ref{eq:fullstate}) ), these constraints limit the space of permissible states to $\dim \hilbert_A = \dim \hilbert_B = 2$, encoding two qubits.
To entangle photons in this encoding using only passive linear optics, the use of ancillas and postselection is necessary \cite{Kok2007b}, so $M_H > 0$.

\subsection{Generating Bell states with three photons is impossible}
\label{sec:threephotons}

\begin{figure}[h]
\includegraphics[width=.45\textwidth]{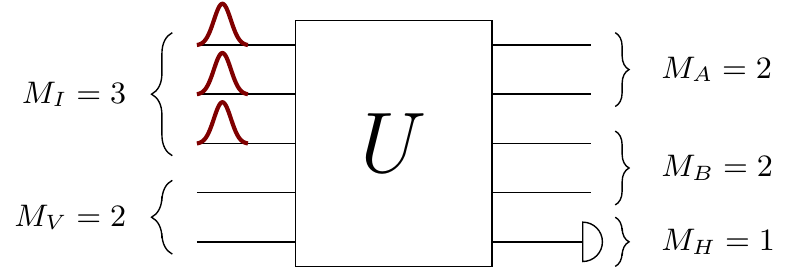}
\caption{
The setup used in Section \ref{sec:threephotons}, with $M_I = n = 3$, $M_V = 2$, $M_A = M_B = 2$, and $M_H = 1$.
We show that no such setup can create an entangled state in dual-rail qubit encoding with any non-zero probability.
On the other hand, with $4$ input photons it is possible to create a Bell state with probability of 1/4 \cite{Zhang2008}.
\label{fig:setup5x5}
}
\end{figure}

It is known that generating a Bell state in dual-rail encoding with just two photons is impossible \cite{Kok2007b, Kieling2008}.
Here we prove that not only is it impossible with three photons, it is only possible to create product states.\par

\begin{proposition}
In a passive linear optical setup using dual-rail encoding, ancillas and postselection, it is not possible to create an entangled state using 3 photon input.
\end{proposition}

\textit{Proof.}
First, let us consider the case where there are five modes ($M =5$); four system modes ($M_A+M_B=4$) and one ancilla ($M_H = 1$), as illustrated in Figure \ref{fig:setup5x5}.
Let the input be three unbunched photons ($n=M_I=3$).
Dual-rail encoding has a total of two photons in a valid qubit state output ($n_S = 2$), implying here that one photon is detected ($n_H=1$).
As there is only one measurement ancilla, the only possible measurement pattern is $\underline{h} = (1)$ (one photon in the fifth mode).\par

As discussed in Sec.~\ref{sec:background}, the amplitudes are related to the permanents of the matrix $U$:
\begin{equation}
\gamma_{kj}((1),U) =
\begin{cases}
 \frac{1}{2} \sum_{\sigma \in S_3} U_{k,\sigma(1)} U_{k,\sigma(2)}  U_{5,\sigma(3)} , & k = j\\
\sum_{\sigma \in S_3} U_{k,\sigma(1)} U_{j,\sigma(2)} U_{5,\sigma(3)} 
&   k \neq j\label{eq:bellcoef}
\end{cases}
\end{equation}
defined $\forall k,j  \in \{1,2,3,4\}$.
The unnormalized state following detection is
\begin{align}
\ket{\widetilde{\psi}((1), U)}
 &= \sqrt{2} \gamma_{11} \ket{2000} + \sqrt{2} \gamma_{22} \ket{0200} \nonumber \\
 & \quad + \sqrt{2} \gamma_{33} \ket{0020} + \sqrt{2} \gamma_{44} \ket{0002} \nonumber \\
 & \quad + \gamma_{13} \ket{1010} + \gamma_{24} \ket{0101} \nonumber \\
 & \quad + \gamma_{12} \ket{1100} + \gamma_{34} \ket{0011} \nonumber \\
 & \quad + \gamma_{14} \ket{1001} + \gamma_{23} \ket{0110} ,
\label{eq:fullstate}
\end{align}
occurring with probability $P((1), U) = 2 \sum_{k=1}^{4} | \gamma_{kk}|^2 + \sum_{\substack{k,j=1 \\ k \neq j}}^{4} | \gamma_{kj} |^2$.\par

In dual-rail encoding it is possible to do any local unitary deterministically by adding beamsplitters and phase shifters to each of the qubits \cite{Kok2007b}.
Thus it suffices to show that it is not possible to create any state of the form $\alpha \ket{0}_A \ket{0}_B + \beta \ket{1}_A \ket{1}_B$ where $| \alpha |^2 + | \beta |^2 = 1$ and $ \alpha \neq 0$, $\beta \neq 0$, because any entangled pure state can be transformed into one of this form by local unitary operations.
The coefficients must therefore satisfy
\begin{align}
\gamma_{11} & = \gamma_{22} = \gamma_{33} = \gamma_{44} = 0, \\
\gamma_{12} & = \gamma_{14} = \gamma_{23} = \gamma_{34} = 0 \quad \mathrm{and} \quad \\
| \gamma_{13} | & = \alpha \sqrt{p}, | \gamma_{24} | = \beta \sqrt{p}, \label{eq:bellcstr}
\end{align}
where $p = P((1),U)$, the probability of one photon being detected in the last mode.
We will now try to find a unitary $U$ that satisfies these constraints.
Define $K_k := U_{k2} U_{53} + U_{k3} U_{52}, \forall k \in \{1, \dots, 4\}$.\par

First, let us consider the case where at least one of $U_{51}$, $U_{52}$ and $U_{53}$ is $0$.
Without loss of generality (wlog) we can label modes so that $U_{51} = 0$, because we can swap $A$ for $B$ and mode $1$ for $2$ without affecting entanglement.
Then the equations in $(\ref{eq:bellcoef})$ can be rewritten as $\gamma_{kk} =  U_{k1} K_{k}$ and $\gamma_{kj} = U_{k1} K_{j} + U_{j1} K_{k}$ for $k \neq j$.
Since $\gamma_{11} = U_{11} K_{1} = 0$ and $\gamma_{13} = U_{11} K_{3} + U_{31} K_{1} \neq 0$, then one and only one of $U_{11}$ or $K_{1}$ can be equal to $0$.
First, assume that $U_{11} = 0$.
Since $K_1 \neq 0$, from the constraints $\gamma_{12} = U_{21} K_{1} = \gamma_{14} = U_{41} K_{1} = 0$ and $\gamma_{24} = U_{21} K_{4} + U_{41} K_{2} \neq 0$, we see that there is no solution.
Similarly, if $K_{1} = 0$, then  $U_{11} \neq 0$ and the constraints $\gamma_{12} = U_{11} K_{2} = \gamma_{14} = U_{11} K_{4} = 0$ and $\gamma_{24} = U_{21} K_{4} + U_{41} K_{2} \neq 0$ again results in no solution.
Therefore there is no solution for which at least one of $U_{51}$, $U_{52}$, $U_{53}$ is zero.\par

Next we assume $K_{k} \neq 0$ $\forall k \in \{1, \dots, 4\}$, with $U_{51} U_{52} U_{53} \neq 0$.
Then solving for $U_{k1}$ from $\gamma_{kk} = 0$ we get $U_{k1} = - U_{k2} U_{k3} U_{51} / K_{k}, \forall k \in \{1, \dots, 4\}$.
Substituting this into the expression for $\gamma_{kj}$ we get 
\begin{equation}
\gamma_{kj} = \frac{U_{51}U_{52}U_{53} (U_{k2}U_{j3} - U_{j2}U_{k3})^2 }{K_k K_j},
\end{equation}
 for all $k,j \in \{1, \dots, 4\}, k \neq j$.
The only way  $\gamma_{12} = \gamma_{23} = 0$, is if $U_{12} U_{23} = U_{22}U_{13}$ and $U_{22} U_{33} = U_{32}U_{23}$.
If $U_{22} U_{23} \neq 0$ then $U_{12} U_{33} = U_{13} U_{32}$, which means $\gamma_{13} = 0$ also, thus cannot be a solution.
If only one of $U_{22}$ or $U_{33}$ is zero, assume $U_{2j} = 0$ where $j$ is $2$ or $3$.
But then $U_{1j} = U_{3j} = 0$ and again $\gamma_{13} = 0$.
If both are zero, then $\gamma_{24} = 0$.
Therefore, there is no solution with $K_{k} \neq 0 \, \forall k \in \{1, \dots, 4\}$.\par

Lastly, assume that at least one of the $K_k = 0$ and that $U_{51} U_{52} U_{53} \neq 0$; wlog, $K_{1} = 0$.
Then $U_{12} = - {U_{13} U_{52}}/{U_{53}}$ combined with the constraint $\gamma_{11} = U_{12} U_{13} U_{51} = 0$ means $U_{12} = U_{13} = 0$.
This gives $\gamma_{1j} = U_{11} K_{j}, \forall j \in  \{1, \dots ,4\}$.
Since $\gamma_{12} = \gamma_{14} = 0$ and $\gamma_{13} \neq 0$, then $U_{11} \neq 0$, while $K_{2} = K_{4} = 0$.
However, this implies $U_{22} = U_{23} = 0$ by a similar argument, further implying that $\gamma_{24} = 0$ and hence there is no solution.\par

We see that under no conditions is there a solution to the given equations where $\alpha \neq 0$ and $\beta \neq 0$.\par

This proves the claim for $5$ modes.
To see that it is true for any number of vacuum ancillas, notice that as long as there are no photons added, Eqs.~(\ref{eq:bellcoef}) do not change other than the mode number $5$ being replaced with the new detection ancilla.
Each new case therefore gives rise to the same constraints implied by Eqs.~(\ref{eq:bellcstr}), with a lack of solution in the same way.
Thus, vacuum ancillas can only increase the probability of creating a state if that probability was nonzero in the first place.\par

Finally, if we allow inputs other than completely unbunched, Eqs.~(\ref{eq:bellcoef}) become even more restrictive.
For example, if there were two photons in input mode $1$ and one photon in input mode $2$, then the matrix elements $U_{i3}$, $U_{3i}$ would not appear in Eqs.~(\ref{eq:bellcoef}), serving only to make the constraints harder to satisfy.\par

$\QED$

\begin{corollary}
In a passive linear optical setup using dual-rail encoding, ancillas and postselection, it is not possible to create a Bell state using 3 photon input\footnote{Kieling observed this using an algebraic geometry approach to the problem \cite{Kieling2008}; here we offer an explicit proof applicable to any entangled state.}.
\end{corollary}

\subsection{Optimal Bell state generation} 
\label{sec:bellstates}

\begin{figure}[h]
    \centering
    \includegraphics[width=.45\textwidth]{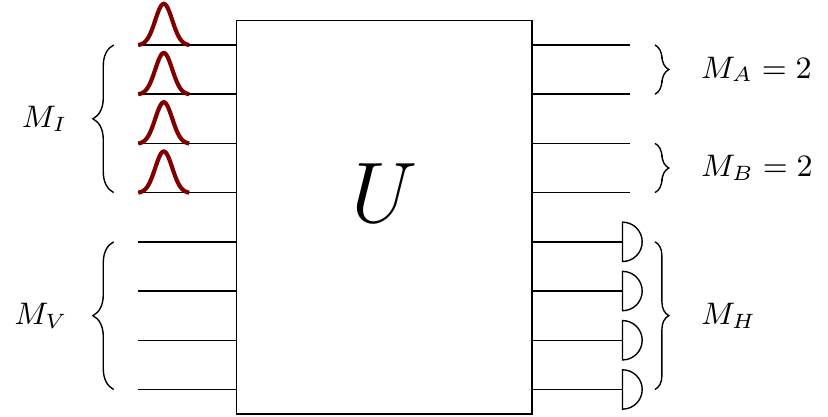}
	\caption{
	The setup used in Section \ref{sec:bellstates} with four photons in eight modes; $M_I = n = 4$, $M_V = 4$, $M_A = M_B = 2$, $M_H = 4$.  
	We give extensive numerical evidence for optimal Bell state generation using this setup when looking for specific Bell states as output.
	}
	\label{fig:setup8x8}
\end{figure}

The previous section showed that Bell state generation with non-zero success probability requires at least four photons. 
Two schemes which accomplish this task using four photons use six~\cite{Carolan2015} and eight~\cite{Zhang2008, Joo2007} modes, with success probabilities of 2/27 and 1/4 respectively.\par

We performed a numerical search for a linear optical Bell state generator that gives a higher success probability.
We used a gradient descent based optimization algorithm over $M = 8$ unitaries with $n = 4$ photon input.
Numerical optimization was carried out in Python, using the BFGS algorithm from the SciPy library \cite{Jones2001}.
This algorithm finds local minima so it needs to be run many times with different seed unitaries, which were randomly selected according to the Haar measure.\par

The cost function we consider is based on the overlap with the desired Bell states.
We allow for six different Bell states, which in the Fock basis after measurement correspond to $\ket{B_{1,2}} = (\ket{1010} \pm \ket{0101}) /  \sqrt{2}$, $\ket{B_{3,4}} =  (\ket{1001} \pm \ket{0110}) / \sqrt{2}$ and $\ket{B_{5,6}} =(\ket{1100} \pm \ket{0011}) / \sqrt{2}$, where the latter can be corrected to the usual dual-rail qubit encoding using a switch \cite{Zhang2008}.
After detecting measurement pattern $\underline{h}$, the overlap between each of these states with the post-selected state is calculated.
We found that raising the overlap to the exponent $10$ optimized the numerical efficacy, penalizing states far from a Bell state heavily.
Multiplying by the probability of detection gives the target cost function to be minimized, $f(\underline{h}, U) = - \sum_{\underline{h}} P(\underline{h}, U)  \sum_{k=1}^{6} | \braket{B_{k} | \psi(\underline{h}, U)} | ^{10}$.\par

\begin{figure}[h]
    	\centering
		\includegraphics[width=.45\textwidth]{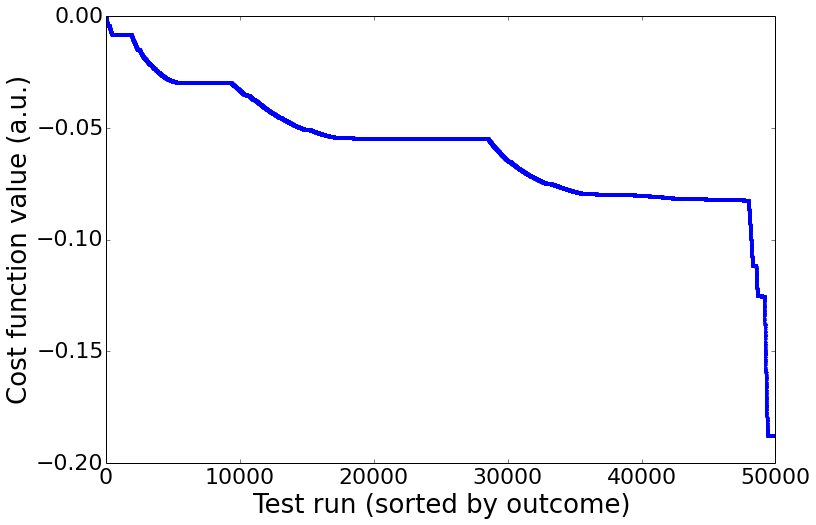}
		\caption{Results of optimization looking for interferometers that generate Bell states with highest probability.
		The minimum found of $\approx - 0.1875$ is exactly bounded by the values of cost function for the known $U_{\text{Bell}}$ interferometer as described in the text.
		Out of $50,000$ test runs, $1.2$~$\%$ of minima found were within $0.0001$ of the minimum corresponding to   $U_{\text{Bell}}$.
		Besides the trials depicted in this graph, the cost function was also optimized with other parameters given to the optimizing algorithm as well as over the space of orthogonal matrices.
		Thus the number of test runs for which a better solution could not be found is close to $100,000$.}
	\label{fig:bellstatenum}
\end{figure}

Figure \ref{fig:bellstatenum} shows the results of this minimization.
The optimal known scheme, when evaluated for this cost function, gives a value of approximately $-0.1875$.
It produces one of these 6 Bell states with probability $1/32$ for $6$ out of the $10$ possible measurement patterns \cite{Zhang2008}.
We can see from the figure that the minimum achieved by the numerical optimization over $50,000$ trials is also approximately $-0.1875$, thus giving solutions which are equivalent to the known scheme in terms of this parameter.
While not a proof, this numerical evidence strongly suggests that the known scheme is optimal for generating the above set of Bell states.
Other cost functions were also attempted, as well as other optimization libraries, but all gave the same results as the technique above. \\

We also investigated the case of non-orthogonal Bell states; for example, allowing $\ket{00} + \ket{11}$ as well as $\ket{00} + i \ket{11}$ as target states.
The possibility of both of these states being generated from the same $U$ for different measurement patterns was explored by running similar numerical optimizations rewarding such situations.
We found no such unitary, which is an interesting result in itself.\\

Though the complexity of the problem grows quickly, we also looked at how the situation changes with higher numbers of input photons and modes.
We numerically optimized over $n=5$, $M=10$ using a similar algorithm and no improved solutions were found over $5000$ runs.
Similarly, we checked $n=6$, $M=12$ over $1000$ runs and here as well there was no improvement over the $-0.1875$ result for $n=4$, $M=8$.

\section{Random unitaries}
\label{sec:comparison}

In this section, we move from the dual-rail qubit encoding of Section~\ref{sec:qubit} to mode entanglement in Section~\ref{sec:modeEntanglement}.
First, we look at how much mode entanglement can be generated with random elements of the unitary group, which we can then use to compare with the dual-rail encoding.
We do so by setting Alice and Bob's number of modes to 2, and numerically computing the average amount of entanglement over measurement patterns.
Notice that this is different from the setting in Section~\ref{sec:qubit}, where we aimed to get a maximally entangled Bell state with the highest possible probability.
Here and in the rest of this work we will study this average entanglement, namely 
\begin{align}
\avgbracket{S(U)}_H  = \sum_{\underline{h} } P(\underline{h}, U) S(\rho_A(\underline{h}, U)) .
\label{eq:avgOverH}
\end{align}
The expectation over the unitary group (for fixed $M$ and $n$) is then $\avgbracket{S}_{H, \mathrm{U}} = \int_{\mathrm{U}(M)} dU\, \avgbracket{S(U)}_H$, where $dU$ is the normalized Haar measure.\par

\label{subsec:numericalavg}

\begin{figure}[h]
\centering
\includegraphics[width=.45\textwidth]{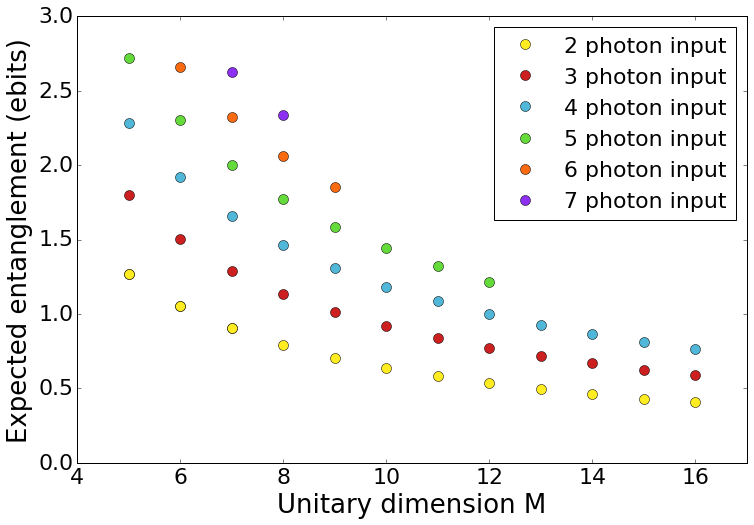}
\caption{
The expectation, over the unitary group, of the average, over measurement patterns, mode entanglement versus the number of modes $M$, for various numbers of unbunched input photons. 
$M_A = M_B = 2$, and if the number of photons $n$ is smaller than $M$, vacuum input modes are added.
The number of heralding detectors is $M_H = M - M_A - M_B$.
The entanglement for a single unitary $U$ is averaged over all measurement patterns, and subsequently averaged over 100,000 randomly Haar-sampled unitaries $U$.
Colours represent different number of input photons, with $2 \leq n \leq 7$.
}
\label{fig:N2average}
\end{figure}

Figure \ref{fig:N2average} shows the numerical results.
We notice that often the average is higher than $1$ ebit, which is the maximum we can achieve in dual-rail qubit encoding.
Adding input photons for the same $M$ increases the average entanglement, while adding vacuum ancillas decreases it.
We see that the average entanglement of $n+1$ photons in $M+1$ modes can be lower than that for $M$ and $n$ (see $n=M=5$ and $6$).
That is, we do not expect more average entanglement by adding a photon at the cost of adding another mode.
Further, we note that even with $2$ photons, there is more average entanglement generated than in the optimal Bell state generator with $4$ photons.
We explore this in more detail for a better comparison.\\

\begin{figure}[h]
\centering
\includegraphics[width=.45\textwidth]{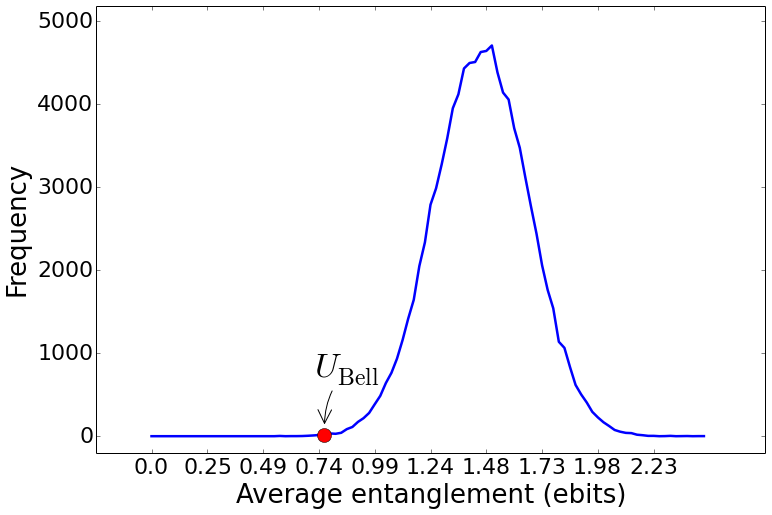}
\caption{
Numerical evaluation of $\avgbracket{S(U)}_H $ for $100,000$ unitaries $U$ chosen using the Haar measure in the case $M_A = M_B = 2$, $M_H = 4$, and $n=4$ unbunched input photons.
Average entanglement for a given $U$ was calculated according to Equation (\ref{eq:avgOverH}) and then binned in one of $100$ bins with a minimum of $0$ and maximum obtained in the samples.
The red dot marks the value of average entanglement that the Bell generating unitary from Section \ref{sec:bellstates} can give, denoted as $U_{\text{Bell}}$, if all of its output states were used.
}
\label{fig:4x4comparison}
\end{figure}

In the usual Bell state generation scenario discussed in Section \ref{sec:bellstates}, if the measurement outcome indicates that the output state is outside of the qubit subspace, the output is discarded.
Here we include the entanglement of the discarded states in accordance with Eq. (\ref{eq:avgOverH}).
We compare the optimal Bell state generator to random unitaries with the same parameters; $M_A = M_B = 2$, $n=4$ and $M=8$.\par

In Fig.~\ref{fig:4x4comparison} we see the results of the comparison.
Firstly, in Section \ref{sec:bellstates} we saw that the probability of getting a Bell state for a state correctable with a single switch is $3/16$ \cite{Zhang2008}. A Bell state gives a single ebit, and if all the other states are discarded, the average entanglement would be $0.1875$ ebits.
If all the outputs from this unitary were counted towards average entanglement as discussed in the previous paragraph (where Equation \ref{eq:avgOverH} is utilized), the entanglement obtained is marked on the Figure \ref{fig:4x4comparison} as $U_{\text{Bell}}$.
As we can see from the graphs, $U_{\text{Bell}}$ gives a markedly lower amount of entanglement than what could be generated on average with a random unitary on the same number of modes.

\section{Mode entanglement}
\label{sec:modeEntanglement}

The previous section shows that, on average, random unitaries give significantly more mode entanglement than dual-rail encoding.
We therefore turn our attention to the investigation of mode rather than qubit entanglement as defined in Section \ref{sec:background}. \par

Equation~(\ref{eq:hilbertns}) states that the total system Hilbert space is a direct sum of Hilbert subspaces such that the sum of Alice and Bob's photon numbers is $n_S$, the number of photons left after heralding.
Let $\rho_{AB} = \ket{{\psi}_{S} (\underline{h}, U)}  \bra{{\psi}_{S} (\underline{h}, U)} $ as in Eq.~(\ref{eq:state}).
Alice's reduced density matrix is 
\begin{align}
\rho_{A} (\underline{h}, U) & = \Tr_B [\rho_{AB} (\underline{h}, U)] \\
& = \sum_{\underline{b}''} \bra{\underline{b}''} \left( 
\sum_{\underline{a},\underline{b},\underline{a}',\underline{b}'} C_{\underline{a} \underline{b}} \overline{C}_{\underline{a}' \underline{b}'} \ket{\underline{a} \underline{b}} \bra{\underline{a}' \underline{b}'} \right) \ket{\underline{b}''} \nonumber \\
& = \sum_{\underline{a}, \underline{a}'} \left(  \sum_{\underline{b}} C_{\underline{a} \underline{b}} \overline{C}_{\underline{a}' \underline{b}} \right) \ket{ \underline{a}} \bra{  \underline{a}'},
\end{align}
where only the terms with $\| \underline{a} \|_1 = \| \underline{a}'\|_1$ are non-zero, because $\| \underline{b}\|_1 = \| \underline{b}'\|_1 = \| \underline{b}''\|_1$ and $n_S=\| \underline{a} \|_1+\|\underline{b} \|_1 = \| \underline{a}'\|_1+\|\underline{b}' \|_1$.
Therefore, there exists a Fock basis ordering in which Alice's reduced state is block diagonal, which allows us to derive a bound on the entanglement (see Section \ref{subsec:dimensionality}).
In the case that Alice has a single mode, this implies her state is diagonal in Fock basis.
The total number of orthogonal states available to Alice is
\begin{equation}
\dim(\hilbert_A^{n_S}) = \sum_{n_A=0}^{n_S} \binom{M_A + n_A - 1}{n_A} = \binom{M_A + n_S}{n_S}.
\end{equation}\par

In Section~\ref{subsec:singlemode}, we find entanglement bounds when Alice only has one mode.
The bound depends on the input state; if the input photons are bunched in a single mode, entanglement is unbounded as the number of photons increases.
Surprisingly, if the input is unbunched, we find a constant bound independent of the number of Bob's modes and independent of the number of photons.
More general bounds can be found, though they are also more loose.
In Section~\ref{subsec:dimensionality} we give the bound on entanglement due to the block diagonal structure of Alice's reduced density matrix in Fock basis. 
In Section~\ref{subsec:linearity} we give a bound which is a consequence of the linearity of the mode transformations.
Unlike in Sec.~\ref{subsec:singlemode}, neither of these bounds depend on the unitarity of the mode transformations, which we expect should affect the amount of entanglement that can be achieved.
In Section~\ref{subsec:unitarity} we conjecture a general unitarity bound based on numerical evidence. 

\subsection{Entanglement when Alice has a single mode}
\label{subsec:singlemode}

\subsubsection{Entanglement for bunched input can be unbounded}
\label{subsec:bunched}

First, we show that mode entanglement is unbounded if we are not restricted to unbunched input.

\begin{figure}[h]
    \centering
    \includegraphics[width=.45\textwidth]{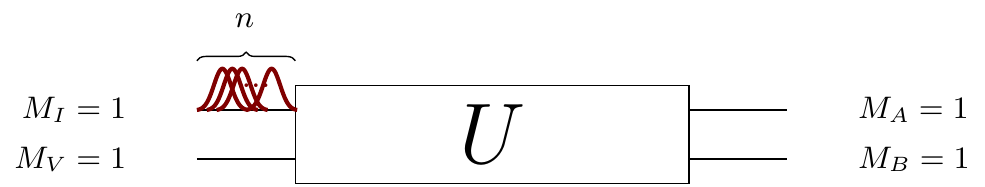}
	\caption{The setup used in Section \ref{subsec:bunched}, where we consider only $M=2$ modes.
	The input consists of all $n$ photons bunched in the top mode; $M_I = M_V = M_A = M_B = 1$, $M_H = 0$.
	We prove that in this setup maximal entanglement grows as $\log{n}$.}
	\label{fig:bunchedsetup}
\end{figure}

\begin{proposition}
Let the input into a $M = 2$ interferometer consist of $n$ photons bunched in a single mode (see Figure \ref{fig:bunchedsetup}).
Then the entanglement across the two output modes is at most $O ( \log {n} )$ ebits, which is achieved when $U$ is a balanced beamsplitter.
\label{prop:bunched}
\end{proposition}

\textit{Proof.}
Parameterize the $M = 2$ unitary matrix $U$ acting on Alice and Bob's single mode Hilbert spaces as
\begin{equation} \label{eq:U2}
U = \left[
\begin{matrix}
c & d \\
-d^{*} & c^{*}
\end{matrix}  \right] ,
\end{equation}
where $|c|^2 + |d|^2 = 1$.
The output state is
\begin{align}
\ket{n0} & = \left( \crop{a}{1} \right)^n / \sqrt{n!} \ket{0} \nonumber \\
& \mapsto \left( c \crop{a}{1} -d^{*} \crop{a}{2} \right)^n / \sqrt{n!} \ket{0} \nonumber \\
& =  \frac{1}{\sqrt{n!}} \sum_{k=0}^{n} \binom{n}{k} (c \crop{a}{1})^k (-d^{*} \crop{a}{2})^{n-k} \ket{0} \nonumber \\
& = \frac{1}{\sqrt{n!}} \sum_{k=0}^{n} \binom{n}{ k} c^k (-d^{*})^{n-k} \sqrt{k!} \sqrt{(n-k)!} \ket{k}\ket{n-k} \nonumber \\
& = \sum_{k=0}^{n} \sqrt{\binom{n}{ k}} c^k (-d^{*})^{n-k} \ket{k}\ket{n-k} .
\end{align}
When Alice has only one mode, her reduced density matrix is diagonal in the Fock basis, so we can find the spectrum of her state directly from the above equations:
\begin{align}
\lambda_k & = \binom{n}{ k} (|c|^2)^k (|d|^2)^{n-k} = \binom{n}{ k} (|c|^2)^k (1-|c|^2)^{n-k} .
\end{align}
This is a binomial distribution with a `success' probability of $p=|c|^2$.
The entropy of the binomial distribution for a fixed $p$ is $1/2 \log_2 \left( 2 \pi e n \cdot p \cdot (1- p) \right)  + O(1 / n)$\footnote{From, e.g., the de Moivre-Laplace Theorem}.
Thus we see that the entanglement bound is $O(\log{n})$, where $n$ is the number of photons.
The constant prefactor is maximized for $p=|c|^2 = |d|^2 = 1/2$, whence the entropy of Alice's state is $1/2 \log_2 \left( 2 \pi e n \cdot 1 / 2 \cdot (1- 1/2) \right) + O(1 / n) = 1/2 \log_2 \left(  \pi e n / 2 \right) + O(1 / n) $.
Finally, notice that solutions to Equation (\ref{eq:U2}) where $|c|^2 = |d|^2 = 1/2$ are a family of balanced beamsplitters.
$\QED$

This is in stark contrast to the situation where the input is unbunched, where we will see in the next section that the entanglement is bounded by a constant.

\subsubsection{Entanglement for unbunched input is bounded}
\label{subsec:oppimConstant}

We now consider situations where Alice only has one mode, Bob can have many, and we do not use any measurement.
The following Lemma will be of use.\par

\begin{figure}[h]
    \centering
    \includegraphics[width=.45\textwidth]{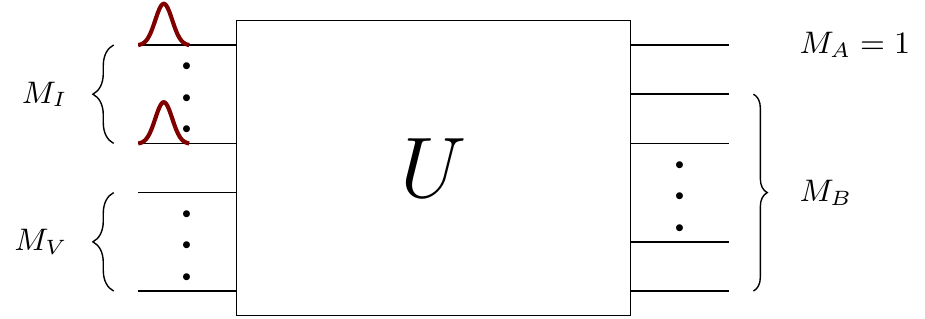}
	\caption{The setup used in Section \ref{subsec:oppimConstant}. 
	The input is an unbunched state with $M_I = n$, with $M_V \geq 0$, $M_A = 1, M_B \geq 1$ and $M_H = 0$.
	We prove that entanglement for this setup is bounded by a constant.}
	\label{fig:setupAlice}
\end{figure}

\begin{lemma}
Consider inputting a Fock state $\ket{\underline{n}} = \ket{n_1  \dots  n_M}$ into an arbitrary interferometer that has $M$ modes.
Let $N = \max{ \{n_1,  \dots ,n_M \} }$.
Then the mean photon number in each output port is bounded by $N$ \cite{Olson2016}.
\label{lemma:meann}
\end{lemma}

\textit{Proof.}
Let $\ket{\underline{n}}$ be an arbitrary Fock state.
\begin{align}
\langle \hat{n}_j \rangle 
 &= \bra{\underline{n}} {\repU^{(n)}}^\dag \hat{n}_j \repU^{(n)} \ket{\underline{n}} \nonumber \\
 &= \bra{\underline{n}} {\repU^{(n)}}^\dag \hat{a}_j^\dag \repU^{(n)} {\repU^{(n)}}^\dag \hat{a}_j \repU^{(n)} \ket{\underline{n}} \nonumber \\
 &= \bra{\underline{n}} \left( \sum_{j'} \hat{a}_{j'}^\dag \overline{U_{j j'}} \right) \left( \sum_{j''} \hat{a}_{j''} {U_{j j''}} \right) \ket{\underline{n}} \nonumber \\
 &= \sum_{j'} \sum_{j''} \overline{U_{j j'}} {U_{j j''}} \bra{\underline{n}} \hat{a}_{j'}^\dag \hat{a}_{j''} \ket{\underline{n}} \nonumber \\
 &= \sum_{j'} \overline{U_{j j'}} {U_{j j'}} \, n_{j'} \label{eq:U}
\end{align}
If, as hypothesized, $n_j \leq N$ for all modes $j$, then

\begin{align}
\langle \hat{n}_j \rangle 
 &= \sum_{j'} | U_{j' j} |^2  \, n_{j'} \leq \sum_{j'} | U_{j' j} |^2  \, N = N .
\end{align}
$\QED$

In the following calculations we shall assume that $n \rightarrow \infty$ as any bound on the entropy found for this infinite case would also hold for a finite one with the same set of constraints.\par

\begin{lemma}
Let $\{p_j\}_{j=0}^{\infty}$ be a probability distribution subject to the constraint $\sum_j j p_j \leq N$.
Then the entropy of this distribution is at most $\log{ \left(  (1 + N)^{1 + N} / N^N \right) }$.
\label{lemma:infinity}
\end{lemma}

\textit{Proof.}
The entropy of the probability distribution $\{p_j\}_{j=0}^{\infty}$ is $S = - \sum_{j=0}^{\infty} p_j \log{p_j}$.
We maximize this subject to the constraints $\sum_{j=0}^{\infty} j p_j = \overline{n} \leq N$ and $\sum_{j=0}^{\infty} p_j = 1$ using the method of Lagrange multipliers.\par

Let the Lagrangian be 
\begin{equation}
L = S + \left( \lambda_0 + \log{e} \right) \left( \sum_{j=0}^{\infty} p_j - 1 \right) + \lambda_1 \left( \sum_{j=0}^{\infty} j p_j -  \overline{n} \right).
\end{equation}
Then $\partial{L} / \partial{p_j} = - \log{p_j} + \lambda_0  + \lambda_1 j$.
Setting $\partial{L} / \partial{p_j} = 0$ gives $p_j = 2^{\lambda_0 + \lambda_1 j}$.
Substituting the value of $p_j$ into the constraints, we get 
\begin{align}
\sum_{j=0}^{\infty} j p_j & = 2^{\lambda_0} 2^{\lambda_1} / (1 - 2^{\lambda_1})^2 =  \overline{n} \\
\sum_{j=0}^{\infty} p_j & = 2^{ \lambda_0}  / (1 - 2^{\lambda_1}) = 1
\end{align}

This allows us to solve for $\lambda_0$ and $\lambda_1$, giving 
\begin{equation}
\lambda_0 = \log{ [ 1 / (1+ \overline{n}) ] }, \,
\lambda_1 =  \log{ [ \overline{n} / (1+ \overline{n}) ] }.
\end{equation}
Notice that 
\begin{align}
S & = -\sum_j p_j \log{p_j} = - \sum_j p_j ( \lambda_0 + \lambda_1 j ) \nonumber \\
& = -\lambda_0 - \lambda_1  \overline{n} \nonumber \\
& = \log{ \left(  \left(1 +  \overline{n} \right)^{1 +  \overline{n}} /  \overline{n}^{\overline{n}} \right) }
\end{align}

The function above increases monotonically for $ \overline{n} \geq 0$ and since $ \overline{n} \leq N$ we get
\begin{equation}
S \leq \log{ \left(  (1 + N)^{1 + N} / N^N \right) }.
\label{eq:entropyBound}
\end{equation}
$\QED$

\begin{corollary}
Let $\{p_j\}_{j=0}^{\infty}$ be some probability distribution subject to the constraint $\sum_j j p_j \leq N$,  $N \in [0,1]$.
Then the entropy of this distribution is at most $2$ ebits.
\label{corr:prob}
\end{corollary}

\begin{theorem}
Let Alice have one output mode, $M_A=1$, and Bob have $M_B = k$.
Let the input be a single photon in each of the $k+1$ modes.
Then the entanglement between Alice and Bob is bounded by $2$ ebits for all $k$.
\label{thm:bound2}
\end{theorem}

\textit{Proof.}
Alice's reduced density matrix is diagonal in the Fock basis, where each entry $\bra{j} \rho_A \ket{j}$ corresponds to the probability that Alice's mode contains $j$ photons.
By Lemma  \ref{lemma:meann}, this distribution satisfies the conditions of Corollary \ref{corr:prob}.
Thus the von Neumann entropy of this state is bounded by $2$ for any $k$, as the bound which holds for $k \rightarrow \infty$ also holds for any finite $k$ as well.
$\QED$

Notice that extra vacuum modes will not increase this limit on the entanglement as the limit is due to the expected number of photons in Alice's mode being at most $1$.
We see that despite the fact that the dimension of Alice's Hilbert space grows with the number of photons as $n + 1$, and Bob's can be even larger, the maximum entanglement is severely constrained to be less than $2$ ebits.\par

Because we are interested in the average entanglement, the result will hold for heralding as well:
\begin{corollary}
Let Alice have one output mode, $M_A=1$, while Bob and Harold have $M_B + M_H = k$.
Let the input be a single photon in each of the $k+1$ modes.
Then the entanglement between Alice and Bob is bounded by $2$ ebits for all $k$.
\label{corr:bound2}
\end{corollary}

\textit{Proof.}
No LOCC operation can increase the amount of entanglement in the system on average \cite{Horodecki2001}.
Therefore, $\avgbracket{S(U)}_H  = \sum_{\underline{h}} P(\underline{h}, U) S(\rho_A(\underline{h}, U)) \leq S(\rho_A(U))$, where $\rho_A(U)$ is Alice's reduced density matrix before any measurement, and by Theorem~\ref{thm:bound2}, $S(\rho_A(U)) \leq 2$ ebits.
$\QED$

We can also examine inputs that have different numbers of bunched photons.
If the highest number of photons in a single input mode is $N$, as per Lemma \ref{lemma:meann}, the expected number of photons in Alice's mode will then be bounded by $N$.
Because the function which bounds the entropy, Eq.~(\ref{eq:entropyBound}), is monotonically increasing, the entropy of Alice's (diagonal) state $( p_0,  \dots  , p_n )$ is at most $\log{ \left(  (1 + N)^{1 + N} / N^N \right) }$ by Lemma~{\ref{lemma:infinity}}.
In the extreme case where all the photons are bunched in a single mode, $S$ scales as $O(\log (N + 1))$, consistent with Proposition~\ref{prop:bunched}.

\subsubsection{Entanglement when Bob also has a single mode}
\label{subsec:oppimMeasurement}

\begin{figure}[h]
    \centering
    \includegraphics[width=.45\textwidth]{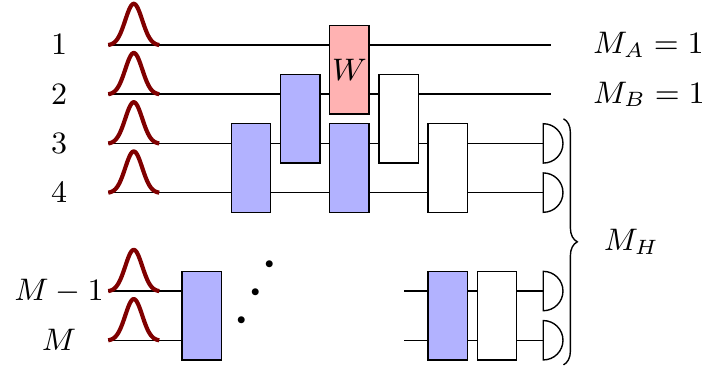}
\caption{
The setup used in Section~\ref{subsec:oppimMeasurement}, where $M = n $, $M_A = M_B = 1$, and $M_H \geq 1$.
An arbitrary $M$ mode interferometer can be decomposed into $M(M+1)/2$ two-mode interferometers~\cite{Hurwitz1897, Reck1994}.
Note that this also applies to an arbitrary $M-1$ mode sub-interferometer (blue).
By focusing on the only component that entangles Alice and Bob (red), we show that the maximum entanglement is the $M=n=2$ value of $\log3$ ebits.
}
\label{fig:setupMeasurementN}
\end{figure}

In this section we consider a similar setup to the previous section, except now we fix the number of Bob's modes to 1 and assign the rest to Harold.
Recall that we are interested in generating the highest amount of entanglement between Alice and Bob on average, thus the probability of detection patterns must be taken into account.
More precisely, we are looking for the maximum of $\avgbracket{S(U)}_H  = \sum_{\underline{h}} P(\underline{h}, U) S(\rho_A(\underline{h}, U))$.
Some patterns might yield a state with high entanglement, but be very unlikely to occur.
In a practical setting we might prefer states that are less entangled but we can generate more consistently.

We first prove a technical lemma that will be useful later.

\begin{lemma}
Given a probability distribution $(p_0, \dots ,p_n)$ such that $\sum_{j=0}^{n} j p_j = 1$, the sum $\sum_{{ j=0}}^{{ n}} p_j \log{(j+2)}$ is bounded by $\log{3}$ which can be achieved by  $p_1 = 1$ and $p_k = 0$ for $k \in \{0,2,3, \dots ,n\}$.
\label{lemma:probMixed}
\end{lemma}

\textit{Proof.}
Since $f(x) = \log{(x+2)}$ is a concave function, by Jensen's inequality $\sum_{{ j=0}}^{{ n}} p_j f(j) \leq f\left( \sum_{{ j=0}}^{{ n}} p_j j \right) = f(1) = \log{3}$, which is achieved by substituting $p_1 = 1$ and $p_k = 0$ for $k \in \{ 0,2,3, \dots ,n \}$.
$\QED$

\begin{theorem}
Consider an interferometer with $M \geq 3$ modes, where both Alice and Bob have one mode and the other output modes are measured using photon counting detectors.
Let the input be the $n = M$ unbunched Fock state.
Then the maximal average entanglement that can be created between Alice and Bob is $\log{3}$ ebits.
\end{theorem}

\textit{Proof.}
First, notice that the average entanglement achievable by an $M = 2 $ interferometer can be achieved for $M \geq 2$ by having modes $3$ to $M$ transform trivially, since photons in these modes will be detected with unit probability.
Thus $\max{\avgbracket{S(U_{ M})}_H} \geq \max{ \avgbracket{S(U_{ M=2})}_H}  = \log{3}$ ebits, $\forall M \geq 3$. 
The interferometer given in Section \ref{subsec:numericalmax}, Eq.~(\ref{eq:beamsplitter2}) below achieves this.

Any $U \in$ U$(M)$ can be decomposed as in Fig.~\ref{fig:setupMeasurementN}.
Then the bottom left triangle (colored blue in the figure) is a unitary $V \in$ U$(M-1)$.
Since the input is unbunched, Lemma~\ref{lemma:meann} implies that each output from $V$ has a mean photon number of $1$.
In particular, Bob's mode before beamsplitter $W$ (red in the figure) will satisfy $\sum_k k q_k=1$ where $k$ is the number of photons occuring with probability $q_k$.
Since the remaining beamsplitters (white in the figure) act only on Bob and Harold's systems, they have no effect on Alice's reduced state and can therefore be ignored.

Let the probability of detecting pattern $\underline{h}$ be $p_{\underline{h}}$, and the probability of detecting a total of $n_H=\|\underline{h}\|_1$ photons be $p_{n_H}=\sum_{\underline{h}:\|\underline{h}\|_1=n_H} p_{\underline{h}}$.
The average entanglement is
\begin{align}
\avgbracket{S(U)}_H 
 &=\sum_{\underline{h}} p_{\underline{h}} S(\rho_{A}(\underline{h})) \nonumber \\
 &=\sum_{n_H = 0}^{n} p_{n_H} \sum_{\underline{h}:\|\underline{h}\|_1=n_H} p_{\underline{h}} / p_{n_H} S(\rho_{A}(\underline{h})) \nonumber \\
 &\leq \sum_{n_H = 0}^{n} p_{n_H} \sum_{\underline{h}:\|\underline{h}\|_1=n_H} p_{\underline{h}} / p_{n_H} \log{(n - n_H + 1)} \nonumber \\
  &=\sum_{n_H = 0}^{n-1} p_{n_H} \log{(n - n_H + 1)} ,
\end{align}
where we've used the fact that the entanglement of $S (\rho_{A}(\underline{h}) )$ is upper bounded by the Schmidt rank $\log(n-n_H+1)$.

As the photon number found in modes $1$ and $2$ is set before the beamsplitter $W$, if $n_H$ photons have been detected, then there were already $n_H$ photons in modes $3$ through $M$.
Alice contributes one photon through her mode to their joint system, which implies that Bob must contribute $n - n_H - 1$ photons through mode $2$, occurring with probability $q_{n - n_H - 1}$.
Therefore $p_{n_H} = q_{n-n_H - 1}$ and recall that Bob's probability distribution is constrained by $\sum_{k=0}^{n-1} k q_k=1$.
By Lemma~\ref{lemma:probMixed} $\sum_{n_H = 0}^{n - 1} q_{n - n_H - 1} \log{(n - n_H + 1)} = \sum_{j = 0}^{n-1} q_{j} \log{(j+2)}$ is maximized for $j = n - n_H - 1 = 1$, that is $q_1 = 1$, yielding $\avgbracket{S(U)}_H  \leq \log{3}$.
This also implies that $n_H  = n - 2$ photons are detected in the optimal situation.
$\QED$

Note that this agrees with the bound in Theorem \ref{thm:bound2} found in the previous section, which follows from the entanglement measure property $\avgbracket{S(U)}_H  = \sum_{\underline{h}} P(\underline{h}, U) S(\rho_A(\underline{h}, U)) \leq S(\rho_A(U))$, where $\rho_A(U)$ is Alice's reduced density matrix before any measurement.
Here the maximum entanglement is $\log 3 < 2$ ebits.
Moreover, adding more vacuum input modes will not affect this bound, as this would only change Bob's expected number of photons before the beamsplitter $W$ to be at most $1$ instead of exactly $1$ as per Lemma \ref{lemma:meann}.

\subsection{Entanglement when Alice has many modes}
\label{sec:DimLin}

In this section we give two bounds on entanglement for more general situations when Alice has more than one mode, based on the Schmidt rank of Alice's reduced state.
They are independent of the input state or any interferometer transformation, depending only on the given number of photons and modes; we assume the latter is the same for both Alice and Bob.
This generality comes at a price, however, in that the bounds loosen; we will discuss a conjectured tighter bound in the following section.

\subsubsection{Dimensionality}
\label{subsec:dimensionality}

By looking solely at the dimensions of Alice and Bob's Hilbert spaces, we can derive an entanglement bound as follows.

\begin{proposition}
Let Alice's and Bob's joint postselected state have a total of $n_S$ photons. 
Let Alice and Bob have $M_A=M_B$ modes.
The Schmidt rank, $\omega$, is at most
\begin{align}
\omega &= 2 \binom{M_A + \frac{n_S - 1}{2}} { \frac{n_S - 1}{2} } & n_S \text{ odd,} \\
\omega &= 2 \frac{n_S + M_A}{n_S}  \binom{M_A + \frac{n_S}{2} - 1}{\frac{n_S}{2} - 1} & n_S \text{ even.}
\end{align}
\label{prop:schmidtRank}
\end{proposition}

\textit{Proof.}
Let Alice's and Bob's joint state be $\ket{\psi_{S} (\underline{h}, U)} = \sum_{k,j} C_{kj} (\underline{h}, U) \ket{k}_A \otimes \ket{j}_B$, where we include the possibility of no measurement ($M_H=0$). 
The Schmidt decomposition is achieved by a state dependent change of basis such that 
\begin{align}
\ket{\psi_{S} (\underline{h}, U)} = \sum_{q = 1}^{\min ( \dim \hilbert_A, \dim \hilbert_B ) } \lambda_{q}  \ket{q}_A \otimes \ket{q}_B ,
\end{align}
where $\{  \ket{q}_{A,B}  \}$ are orthonormal bases for $A$ and $B$, respectively.

Writing this state in terms of Alice and Bob's photon numbers we have $\ket{\psi_{S} (\underline{h}, U)} = \sum_{n_A=0}^{n_S} \ket{\psi_{S}^{n_A, n_B} (\underline{h}, U)}$ with $n_B=n_S-n_A$.
The overlap $\braket{ \psi_{S}^{n_A, n_B} (\underline{h}, U) | \psi_{S}^{n'_A, n'_B} (\underline{h}, U)} = 0$ for $n_A \neq n'_A$, $n_B \neq n'_B$ as these states belong to different Hilbert subspaces in the direct sum.
The reduced density matrix is block diagonal -- each block corresponds to a different $(n_A, n_B)$ combination.
We may therefore consider each subspace individually, where the maximal Schmidt rank is $\min ( \dim \hilbert_A^{n_A}, \dim \hilbert_B^{n_B} )$.
The total number of Schmidt coefficients is therefore at most
\begin{align}
\omega &=  \sum_{n_A = 0}^{n_S} \min \{ \dim \hilbert_A^{n_A}, \dim \hilbert_B^{n_B} \} \\
&= \sum_{n_A = 0}^{n_S} \min \left\{ \binom{M_A + n_A - 1}{ n_A}, \binom{M_B + n_B - 1}{n_B} \right\} .
\end{align}
For $M_A = M_B$ this gives the result.
$\QED$

Since the entanglement is given by the number of nonzero Schmidt coefficients, this gives a bound on the entanglement $S \leq \log(\omega)$.

\subsubsection{Linearity bound}
\label{subsec:linearity}

Here we consider a bound due to the linearity of the interferometer transformations.
In the following we do not assume anything about the form of the input Fock state, nor whether measurement occurs or not.

\begin{proposition}
Given an $n$ photon Fock state as input to a $M$-mode linear optical device, with Alice and Bob having $M_A$ and $M_B$ output modes respectively, the maximal entanglement achievable between Alice and the rest of the modes for any state  is bounded by $n$ ebits.
\label{prop:linearity}
\end{proposition}

\textit{Proof.}
Starting with the arbitrary linear optical mode transformation in Eq.(\ref{eq:alg}), we can group the sum into Alice's modes and the `rest':
\begin{align}
\crop{a}{k} \mapsto \sum_{j=1}^{M} \crop{a}{j} U_{jk}
 &=\sum_{j=1}^{M_A} \crop{a}{j} U_{jk} + \sum_{j=M_A+1}^{M} \crop{a}{j} U_{jk} \nonumber \\
 &=: \hat{A}_k(U) + \hat{R}_k(U) .
\end{align}
The degree one polynomials $\hat{A}_k(U)$, $\hat{R}_k(U)$ in the creation operators are not canonical raising operators, because e.g. $[\hat{A}_k(U),\hat{A}_{k'}(U)] \neq \delta_{kk'}$.
This means that different monomials in $\{ \hat{A}_k(U) \}_k$ do not necessarily give rise to orthogonal states; however, this can only reduce the Schmidt rank of the resulting state.

An arbitrary input Fock state is of the form $\prod_{k=1}^M (\crop{a}{k})^{n_k} / \sqrt{n_k !} \ket{\text{vac}}$, so that the output state is of the form
\begin{align}
\prod_{k=1}^M & \frac{1}{\sqrt{n_k !}} (\hat{A}_k(U) + \hat{R}_k(U))^{n_k} \ket{\text{vac}}
\end{align}
i.e. it is a product of $n$ terms, not all of which are necessarily different.
We can rewrite it as
\begin{align}
\mathcal{N} & \prod_{k = 1 }^{n} (\hat{A}_{j_k}(U) + \hat{R}_{j_k}(U)) \ket{\text{vac}},
\end{align}
where $j_k \in \{ 1, \dots ,M \}$ and $\mathcal{N}$ is the necessary normalization.
The highest number of monomial terms in this product is bounded by $2^{n}$ and after tracing out Bob and Harold this also bounds the number of monomial terms that can be in Alice's reduced state.
$\QED$

Consider a balanced 50:50 beam splitter coupling one of Alice's modes (say $k$) to one of Bob's modes (say $k + M_A$).
If Alice's mode contained one input photon and Bob's none, we get $1$ ebit of entanglement.
Proposition \ref{prop:linearity} tells us we can only get up to $n$ ebits using $n$ photons, so as long as $n \leq M_A = M_B$, a beamsplitter coupling mode $k$ with mode $k+M_A$ for $k = 1$ through $n$ in this way would give us a state that achieves the bound.

The dimensionality (Section \ref{subsec:dimensionality}) and linearity bounds above hold for all $M$ and all $n$.
We can find numerically the photon number $n_L (M_A) \in  \mathbb{N}$, which depends on the number of Alice's modes.
For a given $M_A$ it represents the number of photons up to which the linearity bound is smaller than the dimensionality bound.
For $n > n_L(M)$, the dimensionality bound is a tighter limit on the entanglement (see Figure~\ref{fig:avg_ent}).

\subsection{Hints of another bound}
\label{subsec:unitarity}

\begin{figure}[h]
    	\centering
		\includegraphics[width=.45\textwidth]{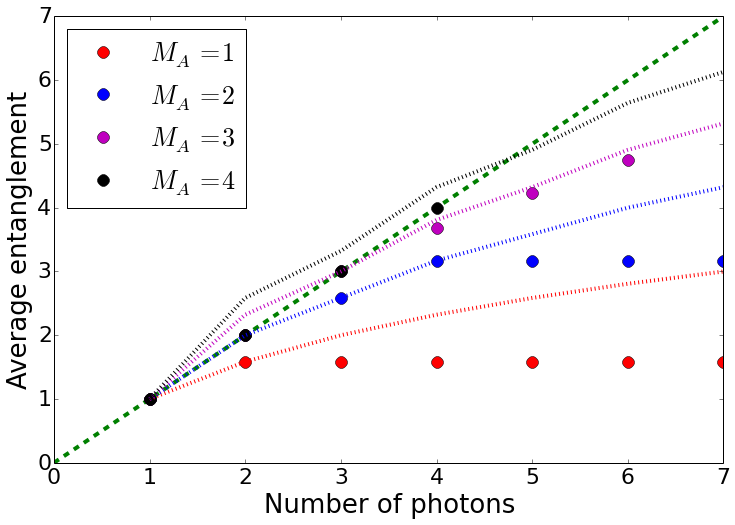}
		\caption{Plot of the maximum average entanglement found through numerical optimization, along with the dimensionality and linearity bounds for $M_A = M_B$. 
		The input are unbunched states.
		If $n > M_A + M_B$, the remaining $M_H = n - M_A - M_B$ modes contain detectors.
		The green dashed line is the linearity bound.
		The dotted lines are dimensionality bounds for the value of $M_A$ whose dots have the same colour.
		We can see that the values of $n_L$ for specific $M_A$s are: $n_L (1) = 1, n_L (2) = 2, n_L(3) = 3, n_L(4) = 4$.
		The dots are solutions returned by the optimization.
}
	\label{fig:avg_ent}
\end{figure}

In this section we explore a potential bound that is motivated by numerical evidence (see Figure \ref{fig:avg_ent}).
While adding more photons to the interferometer increases the size of Alice's and Bob's Hilbert spaces, and according to the results from the previous section should allow for higher amounts of entanglement, we see that this is not what happens in general (assuming the number of modes that Alice and Bob have are fixed).
Based on the analytical results from Section \ref{subsec:singlemode} and the  numerical evidence for all cases up to $n=7$ photons and $M_A = M_B = 3$ modes, we make a conjecture that there is another bound which seems to arise from the unitarity of the mode transformation.

\begin{conjecture}
For $n$ unbunched photons input into an interferometer with $M_A$ Alice and $M_B$ Bob output modes with  $n > M_A + M_B$, the average amount of entanglement, obtained over Harold's measurements, is bounded above by the maximal average amount of entanglement achieved when $n = M_A + M_B$.
\label{conj:unitarity}
\end{conjecture}

We provide numerical evidence supporting this ``unitarity bound'' for various numbers of input photons and modes.
We assume that the input states are unbunched, ancillas and measurement are allowed, and Alice and Bob have the same number of modes; $M_I = n$, $M_V \geq 0$, $M_A = M_B \geq 1$ and $M_H \geq 0$.

Propositions \ref{prop:schmidtRank} and \ref{prop:linearity} provide tight entanglement bounds when all input photons are kept in the system, i.e. when there is no detection.
We know that it is possible to postselect states that exceed these bounds, but because we are interested in average entanglement these cases must be weighted with their heralding probabilities.
Our findings are consistent with a generic trade off between these two quantities, leading to a bounded average entanglement.

Figure \ref{fig:avg_ent} shows the results of numerical optimization of the average entanglement given by Eq.~(\ref{eq:avgOverH}) for various numbers of input photons and modes.
We can see how the linearity and dimensionality bounds of Sec.~\ref{sec:DimLin} are indeed limiting the entanglement.
We also see the appearance of what looks like a third bound, seemingly when the number of photons is larger than the total number of modes in the system ($M_A + M_B$).
This new behaviour is not captured by the bounds we have obtained and we conjecture that it is due to the unitarity of the interferometric transformation.
This leads to the hypothesis that the maximum possible average entanglement, in situations with unbunched input and Alice and Bob have the same number of modes, can be reached using a $(M_A + M_B)$-mode interferometer with $M_A + M_B$ photons.


\subsection{Optimal interferometers}
\label{subsec:numericalmax}

Finally, in this section we report some of the explicit interferometers (unitaries) that produce the optimal entanglement found for small number of modes.

In the case of $M_A = M_B = 1$ and a single photon $n=1$, the well known balanced 50:50 beamsplitter is optimal,
\begin{equation}
BS_{1} = \frac{1}{\sqrt{2}} \left[
\begin{array}{cc}
1&  1 \\
-1& 1
\end{array}
\right] .
\end{equation}
This is familiar, as in single-rail encoding it creates a Bell state.
Let $ \theta = \frac{1}{2} \arccos{(1 / \sqrt{3})} $.
When we input two photons, $n=2$, with one in each mode, the unitary
\begin{equation}
BS_{2} = \left[
\begin{array}{cc}
\cos{\theta}  &  \sin{\theta} \\
-\sin{\theta} &  \cos{\theta}
\end{array}
\right]
\label{eq:beamsplitter2}
\end{equation}
produces a state with $\log{3}$ ebits of entanglement.
Conjecture \ref{conj:unitarity} says that for all higher photon numbers, $\log{3}$ will still be the maximum, achieved by using $BS_{2}$ between any pair of Alice and Bob's modes and identity on all the others (they are just routed straight to the detectors).

For $M_A = M_B = 2$, we have that all the optimal interferometers are actually combinations of $BS_{1}$ and $BS_{2}$. 
An example for $M=n=4$ is:
\begin{equation}
\left[
\begin{array}{cccc}
\cos{\theta}  &  0            &  0            &  \sin{\theta}  \\
0             & \cos{\theta}  &  \sin{\theta} &  0   		   \\
0             & -\sin{\theta} &  \cos{\theta} &  0 			   \\
-\sin{\theta} &  0            &  0            &  \cos{\theta}
\end{array}
\right] ,
\end{equation}
where as before $ \theta = \frac{1}{2} \arccos{(1 / \sqrt{3})} $.
This interferometer corresponds to a $BS_2$ beamsplitter between modes $1$ and $4$ and another $BS_2$ beamsplitter between modes $2$ and $3$, giving $\log{9} \approx 3.17$ ebits of entanglement.
When $n=3$, the optimal value of $\log{6} \approx 2.58$ ebits is achieved by using $BS_2$ on modes $1$ and $4$ and $BS_1$ on modes $2$ and $3$.
For $n=2$, the maximum of $2$ ebits is achieved by two $BS_1$ beamsplitters, similar to $n=4$ case.
Finally, for $n=1$ we just use a single $BS_1$ to achieve $1$ ebit.

\section*{Acknowledgments}
SS was supported by the Bristol Quantum Engineering Centre for Doctoral Training, EPSRC grant EP/L015730/1 and would like to thank P Birchall for bringing to our attention a more elegant proof of Lemma \ref{lemma:meann}.
AM was supported by EPSRC Early Career Fellowship EP/L021005/1.
PST was supported in part by the U.S. Army Research Office under contract W911NF-14-1-0133 and EPSRC First Grant EP/N014812/1.

\bibliography{manual}

\begin{thebibliography}{22}%
\makeatletter
\providecommand \@ifxundefined [1]{%
 \@ifx{#1\undefined}
}%
\providecommand \@ifnum [1]{%
 \ifnum #1\expandafter \@firstoftwo
 \else \expandafter \@secondoftwo
 \fi
}%
\providecommand \@ifx [1]{%
 \ifx #1\expandafter \@firstoftwo
 \else \expandafter \@secondoftwo
 \fi
}%
\providecommand \natexlab [1]{#1}%
\providecommand \enquote  [1]{``#1''}%
\providecommand \bibnamefont  [1]{#1}%
\providecommand \bibfnamefont [1]{#1}%
\providecommand \citenamefont [1]{#1}%
\providecommand \href@noop [0]{\@secondoftwo}%
\providecommand \href [0]{\begingroup \@sanitize@url \@href}%
\providecommand \@href[1]{\@@startlink{#1}\@@href}%
\providecommand \@@href[1]{\endgroup#1\@@endlink}%
\providecommand \@sanitize@url [0]{\catcode `\\12\catcode `\$12\catcode
  `\&12\catcode `\#12\catcode `\^12\catcode `\_12\catcode `\%12\relax}%
\providecommand \@@startlink[1]{}%
\providecommand \@@endlink[0]{}%
\providecommand \url  [0]{\begingroup\@sanitize@url \@url }%
\providecommand \@url [1]{\endgroup\@href {#1}{\urlprefix }}%
\providecommand \urlprefix  [0]{URL }%
\providecommand \Eprint [0]{\href }%
\providecommand \doibase [0]{http://dx.doi.org/}%
\providecommand \selectlanguage [0]{\@gobble}%
\providecommand \bibinfo  [0]{\@secondoftwo}%
\providecommand \bibfield  [0]{\@secondoftwo}%
\providecommand \translation [1]{[#1]}%
\providecommand \BibitemOpen [0]{}%
\providecommand \bibitemStop [0]{}%
\providecommand \bibitemNoStop [0]{.\EOS\space}%
\providecommand \EOS [0]{\spacefactor3000\relax}%
\providecommand \BibitemShut  [1]{\csname bibitem#1\endcsname}%
\let\auto@bib@innerbib\@empty
\bibitem [{\citenamefont {O'Brien}\ \emph {et~al.}(2010)\citenamefont
  {O'Brien}, \citenamefont {Furusawa},\ and\ \citenamefont
  {Vu{\v{c}}kovi{\'{c}}}}]{OBrien2010}%
  \BibitemOpen
  \bibfield  {author} {\bibinfo {author} {\bibfnamefont {J.~L.}\ \bibnamefont
  {O'Brien}}, \bibinfo {author} {\bibfnamefont {A.}~\bibnamefont {Furusawa}}, \
  and\ \bibinfo {author} {\bibfnamefont {J.}~\bibnamefont
  {Vu{\v{c}}kovi{\'{c}}}},\ }\href {\doibase 10.1038/nphoton.2009.229}
  {\bibfield  {journal} {\bibinfo  {journal} {Nature Photonics}\ }\textbf
  {\bibinfo {volume} {3}},\ \bibinfo {pages} {687} (\bibinfo {year}
  {2010})}\BibitemShut {NoStop}%
\bibitem [{\citenamefont {Saffman}\ \emph {et~al.}(2010)\citenamefont
  {Saffman}, \citenamefont {Walker},\ and\ \citenamefont
  {M{\o}lmer}}]{Saffman2010}%
  \BibitemOpen
  \bibfield  {author} {\bibinfo {author} {\bibfnamefont {M.}~\bibnamefont
  {Saffman}}, \bibinfo {author} {\bibfnamefont {T.~G.}\ \bibnamefont {Walker}},
  \ and\ \bibinfo {author} {\bibfnamefont {K.}~\bibnamefont {M{\o}lmer}},\
  }\href {\doibase 10.1103/RevModPhys.82.2313} {\bibfield  {journal} {\bibinfo
  {journal} {Reviews of Modern Physics}\ }\textbf {\bibinfo {volume} {82}},\
  \bibinfo {pages} {2313} (\bibinfo {year} {2010})}\BibitemShut {NoStop}%
\bibitem [{\citenamefont {Xiang}\ \emph {et~al.}(2013)\citenamefont {Xiang},
  \citenamefont {Ashhab}, \citenamefont {You},\ and\ \citenamefont
  {Nori}}]{Xiang2013}%
  \BibitemOpen
  \bibfield  {author} {\bibinfo {author} {\bibfnamefont {Z.-l.}\ \bibnamefont
  {Xiang}}, \bibinfo {author} {\bibfnamefont {S.}~\bibnamefont {Ashhab}},
  \bibinfo {author} {\bibfnamefont {J.~Q.}\ \bibnamefont {You}}, \ and\
  \bibinfo {author} {\bibfnamefont {F.}~\bibnamefont {Nori}},\ }\href {\doibase
  10.1103/RevModPhys.85.623} {\bibfield  {journal} {\bibinfo  {journal}
  {Reviews of Modern Physics}\ }\textbf {\bibinfo {volume} {85}},\ \bibinfo
  {pages} {623} (\bibinfo {year} {2013})}\BibitemShut {NoStop}%
\bibitem [{\citenamefont {Kok}\ \emph {et~al.}(2007)\citenamefont {Kok},
  \citenamefont {Munro}, \citenamefont {Nemoto}, \citenamefont {Ralph},
  \citenamefont {Dowling},\ and\ \citenamefont {Milburn}}]{Kok2007b}%
  \BibitemOpen
  \bibfield  {author} {\bibinfo {author} {\bibfnamefont {P.}~\bibnamefont
  {Kok}}, \bibinfo {author} {\bibfnamefont {W.~J.}\ \bibnamefont {Munro}},
  \bibinfo {author} {\bibfnamefont {K.}~\bibnamefont {Nemoto}}, \bibinfo
  {author} {\bibfnamefont {T.~C.}\ \bibnamefont {Ralph}}, \bibinfo {author}
  {\bibfnamefont {J.~P.}\ \bibnamefont {Dowling}}, \ and\ \bibinfo {author}
  {\bibfnamefont {G.~J.}\ \bibnamefont {Milburn}},\ }\href {\doibase
  10.1103/RevModPhys.79.135} {\bibfield  {journal} {\bibinfo  {journal}
  {Reviews of Modern Physics}\ }\textbf {\bibinfo {volume} {79}},\ \bibinfo
  {pages} {135} (\bibinfo {year} {2007})}\BibitemShut {NoStop}%
\bibitem [{\citenamefont {Jozsa}\ and\ \citenamefont
  {Linden}(2003)}]{Jozsa2003}%
  \BibitemOpen
  \bibfield  {author} {\bibinfo {author} {\bibfnamefont {R.}~\bibnamefont
  {Jozsa}}\ and\ \bibinfo {author} {\bibfnamefont {N.}~\bibnamefont {Linden}},\
  }\href {\doibase 10.1098/rspa.2002.1097} {\bibfield  {journal} {\bibinfo
  {journal} {Proceedings of the Royal Society A: Mathematical, Physical and
  Engineering Sciences}\ }\textbf {\bibinfo {volume} {459}},\ \bibinfo {pages}
  {2011} (\bibinfo {year} {2003})}\BibitemShut {NoStop}%
\bibitem [{\citenamefont {Horodecki}\ \emph {et~al.}(2009)\citenamefont
  {Horodecki}, \citenamefont {Horodecki}, \citenamefont {Horodecki},\ and\
  \citenamefont {Horodecki}}]{Horodecki2009}%
  \BibitemOpen
  \bibfield  {author} {\bibinfo {author} {\bibfnamefont {R.}~\bibnamefont
  {Horodecki}}, \bibinfo {author} {\bibfnamefont {P.}~\bibnamefont
  {Horodecki}}, \bibinfo {author} {\bibfnamefont {M.}~\bibnamefont
  {Horodecki}}, \ and\ \bibinfo {author} {\bibfnamefont {K.}~\bibnamefont
  {Horodecki}},\ }\href {\doibase 10.1103/RevModPhys.81.865} {\bibfield
  {journal} {\bibinfo  {journal} {Reviews of Modern Physics}\ }\textbf
  {\bibinfo {volume} {81}},\ \bibinfo {pages} {865} (\bibinfo {year}
  {2009})}\BibitemShut {NoStop}%
\bibitem [{\citenamefont {Bell}(1964)}]{Bell1964}%
  \BibitemOpen
  \bibfield  {author} {\bibinfo {author} {\bibfnamefont {J.~S.}\ \bibnamefont
  {Bell}},\ }\href@noop {} {\bibfield  {journal} {\bibinfo  {journal}
  {Physics}\ }\textbf {\bibinfo {volume} {1}},\ \bibinfo {pages} {195}
  (\bibinfo {year} {1964})}\BibitemShut {NoStop}%
\bibitem [{\citenamefont {Aspect}(1975)}]{Clauser1969}%
  \BibitemOpen
  \bibfield  {author} {\bibinfo {author} {\bibfnamefont {A.}~\bibnamefont
  {Aspect}},\ }\href {\doibase 10.1016/0375-9601(75)90831-2} {\bibfield
  {journal} {\bibinfo  {journal} {Physics Letters A}\ }\textbf {\bibinfo
  {volume} {54}},\ \bibinfo {pages} {117} (\bibinfo {year} {1975})}\BibitemShut
  {NoStop}%
\bibitem [{\citenamefont {Popescu}\ and\ \citenamefont
  {Rohrlich}(1994)}]{Popescu1994}%
  \BibitemOpen
  \bibfield  {author} {\bibinfo {author} {\bibfnamefont {S.}~\bibnamefont
  {Popescu}}\ and\ \bibinfo {author} {\bibfnamefont {D.}~\bibnamefont
  {Rohrlich}},\ }\href {\doibase 10.1007/BF02058098} {\bibfield  {journal}
  {\bibinfo  {journal} {Foundations of Physics}\ }\textbf {\bibinfo {volume}
  {24}},\ \bibinfo {pages} {379} (\bibinfo {year} {1994})}\BibitemShut
  {NoStop}%
\bibitem [{\citenamefont {Carolan}\ \emph {et~al.}(2015)\citenamefont
  {Carolan}, \citenamefont {Harrold}, \citenamefont {Sparrow}, \citenamefont
  {Martin-Lopez}, \citenamefont {Russell}, \citenamefont {Silverstone},
  \citenamefont {Shadbolt}, \citenamefont {Matsuda}, \citenamefont {Oguma},
  \citenamefont {Itoh}, \citenamefont {Marshall}, \citenamefont {Thompson},
  \citenamefont {Matthews}, \citenamefont {Hashimoto}, \citenamefont
  {O'Brien},\ and\ \citenamefont {Laing}}]{Carolan2015}%
  \BibitemOpen
  \bibfield  {author} {\bibinfo {author} {\bibfnamefont {J.}~\bibnamefont
  {Carolan}}, \bibinfo {author} {\bibfnamefont {C.}~\bibnamefont {Harrold}},
  \bibinfo {author} {\bibfnamefont {C.}~\bibnamefont {Sparrow}}, \bibinfo
  {author} {\bibfnamefont {E.}~\bibnamefont {Martin-Lopez}}, \bibinfo {author}
  {\bibfnamefont {N.~J.}\ \bibnamefont {Russell}}, \bibinfo {author}
  {\bibfnamefont {J.~W.}\ \bibnamefont {Silverstone}}, \bibinfo {author}
  {\bibfnamefont {P.~J.}\ \bibnamefont {Shadbolt}}, \bibinfo {author}
  {\bibfnamefont {N.}~\bibnamefont {Matsuda}}, \bibinfo {author} {\bibfnamefont
  {M.}~\bibnamefont {Oguma}}, \bibinfo {author} {\bibfnamefont
  {M.}~\bibnamefont {Itoh}}, \bibinfo {author} {\bibfnamefont {G.~D.}\
  \bibnamefont {Marshall}}, \bibinfo {author} {\bibfnamefont {M.~G.}\
  \bibnamefont {Thompson}}, \bibinfo {author} {\bibfnamefont {J.~C.~F.}\
  \bibnamefont {Matthews}}, \bibinfo {author} {\bibfnamefont {T.}~\bibnamefont
  {Hashimoto}}, \bibinfo {author} {\bibfnamefont {J.~L.}\ \bibnamefont
  {O'Brien}}, \ and\ \bibinfo {author} {\bibfnamefont {A.}~\bibnamefont
  {Laing}},\ }\href {\doibase 10.1126/science.aab3642} {\bibfield  {journal}
  {\bibinfo  {journal} {Science}\ }\textbf {\bibinfo {volume} {349}},\ \bibinfo
  {pages} {711} (\bibinfo {year} {2015})}\BibitemShut {NoStop}%
\bibitem [{\citenamefont {Gimeno-Segovia}\ \emph {et~al.}(2015)\citenamefont
  {Gimeno-Segovia}, \citenamefont {Shadbolt}, \citenamefont {Browne},\ and\
  \citenamefont {Rudolph}}]{Gimeno-Segovia2015}%
  \BibitemOpen
  \bibfield  {author} {\bibinfo {author} {\bibfnamefont {M.}~\bibnamefont
  {Gimeno-Segovia}}, \bibinfo {author} {\bibfnamefont {P.}~\bibnamefont
  {Shadbolt}}, \bibinfo {author} {\bibfnamefont {D.~E.}\ \bibnamefont
  {Browne}}, \ and\ \bibinfo {author} {\bibfnamefont {T.}~\bibnamefont
  {Rudolph}},\ }\href {\doibase 10.1103/PhysRevLett.115.020502} {\bibfield
  {journal} {\bibinfo  {journal} {Physical Review Letters}\ }\textbf {\bibinfo
  {volume} {115}},\ \bibinfo {pages} {020502} (\bibinfo {year}
  {2015})}\BibitemShut {NoStop}%
\bibitem [{\citenamefont {Joo}\ \emph {et~al.}(2007)\citenamefont {Joo},
  \citenamefont {Knight}, \citenamefont {O'Brien},\ and\ \citenamefont
  {Rudolph}}]{Joo2007}%
  \BibitemOpen
  \bibfield  {author} {\bibinfo {author} {\bibfnamefont {J.}~\bibnamefont
  {Joo}}, \bibinfo {author} {\bibfnamefont {P.~L.}\ \bibnamefont {Knight}},
  \bibinfo {author} {\bibfnamefont {J.~L.}\ \bibnamefont {O'Brien}}, \ and\
  \bibinfo {author} {\bibfnamefont {T.}~\bibnamefont {Rudolph}},\ }\href
  {\doibase 10.1103/PhysRevA.76.052326} {\bibfield  {journal} {\bibinfo
  {journal} {Physical Review A}\ }\textbf {\bibinfo {volume} {76}},\ \bibinfo
  {pages} {052326} (\bibinfo {year} {2007})}\BibitemShut {NoStop}%
\bibitem [{\citenamefont {Zhang}\ \emph {et~al.}(2008)\citenamefont {Zhang},
  \citenamefont {Bao}, \citenamefont {Lu}, \citenamefont {Zhou}, \citenamefont
  {Yang}, \citenamefont {Rudolph},\ and\ \citenamefont {Pan}}]{Zhang2008}%
  \BibitemOpen
  \bibfield  {author} {\bibinfo {author} {\bibfnamefont {Q.}~\bibnamefont
  {Zhang}}, \bibinfo {author} {\bibfnamefont {X.-H.}\ \bibnamefont {Bao}},
  \bibinfo {author} {\bibfnamefont {C.-Y.}\ \bibnamefont {Lu}}, \bibinfo
  {author} {\bibfnamefont {X.-Q.}\ \bibnamefont {Zhou}}, \bibinfo {author}
  {\bibfnamefont {T.}~\bibnamefont {Yang}}, \bibinfo {author} {\bibfnamefont
  {T.}~\bibnamefont {Rudolph}}, \ and\ \bibinfo {author} {\bibfnamefont
  {J.-W.}\ \bibnamefont {Pan}},\ }\href {\doibase 10.1103/PhysRevA.77.062316}
  {\bibfield  {journal} {\bibinfo  {journal} {Physical Review A}\ }\textbf
  {\bibinfo {volume} {77}},\ \bibinfo {pages} {062316} (\bibinfo {year}
  {2008})}\BibitemShut {NoStop}%
\bibitem [{\citenamefont {Rudolph}(2016)}]{Rudolph2016}%
  \BibitemOpen
  \bibfield  {author} {\bibinfo {author} {\bibfnamefont {T.}~\bibnamefont
  {Rudolph}},\ }\href@noop {} {\  (\bibinfo {year} {2016})},\ \Eprint
  {http://arxiv.org/abs/1610.07128} {arXiv:1610.07128} \BibitemShut {NoStop}%
\bibitem [{\citenamefont {Aaronson}\ and\ \citenamefont
  {Arkhipov}(2013)}]{Aaronson2010}%
  \BibitemOpen
  \bibfield  {author} {\bibinfo {author} {\bibfnamefont {S.}~\bibnamefont
  {Aaronson}}\ and\ \bibinfo {author} {\bibfnamefont {A.}~\bibnamefont
  {Arkhipov}},\ }\href@noop {} {\bibfield  {journal} {\bibinfo  {journal}
  {Theory of Computing}\ }\textbf {\bibinfo {volume} {9}},\ \bibinfo {pages}
  {143} (\bibinfo {year} {2013})}\BibitemShut {NoStop}%
\bibitem [{\citenamefont {Scheel}(2004)}]{Scheel2004}%
  \BibitemOpen
  \bibfield  {author} {\bibinfo {author} {\bibfnamefont {S.}~\bibnamefont
  {Scheel}},\ }\href@noop {} {\  (\bibinfo {year} {2004})},\ \Eprint
  {http://arxiv.org/abs/quant-ph/0406127} {arXiv:quant-ph/0406127} \BibitemShut
  {NoStop}%
\bibitem [{\citenamefont {Kieling}(2008)}]{Kieling2008}%
  \BibitemOpen
  \bibfield  {author} {\bibinfo {author} {\bibfnamefont {K.}~\bibnamefont
  {Kieling}},\ }\emph {\bibinfo {title} {{Linear optics quantum
  computing--construction of small networks and asymptotic scaling}}},\ \href
  {http://research.microsoft.com/pubs/72356/Konrad Kieling - Linear optics
  quantum computing.pdf} {Ph.D. thesis},\ \bibinfo  {school} {Imperial College
  London} (\bibinfo {year} {2008})\BibitemShut {NoStop}%
\bibitem [{\citenamefont {Jones}\ \emph {et~al.}(2017)\citenamefont {Jones},
  \citenamefont {Oliphant}, \citenamefont {Peterson} \emph
  {et~al.}}]{Jones2001}%
  \BibitemOpen
  \bibfield  {author} {\bibinfo {author} {\bibfnamefont {E.}~\bibnamefont
  {Jones}}, \bibinfo {author} {\bibfnamefont {T.}~\bibnamefont {Oliphant}},
  \bibinfo {author} {\bibfnamefont {P.}~\bibnamefont {Peterson}},  \emph
  {et~al.},\ }\href {http://www.scipy.org/} {\enquote {\bibinfo {title}
  {{SciPy}: Open source scientific tools for {Python}},}\ } (\bibinfo {year}
  {2001--2017})\BibitemShut {NoStop}%
\bibitem [{\citenamefont {Olson}\ \emph {et~al.}(2016)\citenamefont {Olson},
  \citenamefont {Motes}, \citenamefont {Birchall}, \citenamefont {Studer},
  \citenamefont {LaBorde}, \citenamefont {Moulder}, \citenamefont {Rohde},\
  and\ \citenamefont {Dowling}}]{Olson2016}%
  \BibitemOpen
  \bibfield  {author} {\bibinfo {author} {\bibfnamefont {J.~P.}\ \bibnamefont
  {Olson}}, \bibinfo {author} {\bibfnamefont {K.~R.}\ \bibnamefont {Motes}},
  \bibinfo {author} {\bibfnamefont {P.~M.}\ \bibnamefont {Birchall}}, \bibinfo
  {author} {\bibfnamefont {N.~M.}\ \bibnamefont {Studer}}, \bibinfo {author}
  {\bibfnamefont {M.}~\bibnamefont {LaBorde}}, \bibinfo {author} {\bibfnamefont
  {T.}~\bibnamefont {Moulder}}, \bibinfo {author} {\bibfnamefont {P.~P.}\
  \bibnamefont {Rohde}}, \ and\ \bibinfo {author} {\bibfnamefont {J.~P.}\
  \bibnamefont {Dowling}},\ }\href {http://arxiv.org/abs/1610.07128} {\
  (\bibinfo {year} {2016})},\ \Eprint {http://arxiv.org/abs/1610.07128}
  {arXiv:1610.07128} \BibitemShut {NoStop}%
\bibitem [{\citenamefont {Horodecki}(2001)}]{Horodecki2001}%
  \BibitemOpen
  \bibfield  {author} {\bibinfo {author} {\bibfnamefont {M.}~\bibnamefont
  {Horodecki}},\ }\href@noop {} {\bibfield  {journal} {\bibinfo  {journal}
  {Quantum Information {\&} Computation}\ }\textbf {\bibinfo {volume} {1}},\
  \bibinfo {pages} {3} (\bibinfo {year} {2001})}\BibitemShut {NoStop}%
\bibitem [{\citenamefont {Hurwitz}(1897)}]{Hurwitz1897}%
  \BibitemOpen
  \bibfield  {author} {\bibinfo {author} {\bibfnamefont {A.}~\bibnamefont
  {Hurwitz}},\ }\href@noop {} {\bibfield  {journal} {\bibinfo  {journal}
  {Nachrichten von der Gesellschaft der Wissenschaften zu G{\"o}ttingen,
  Mathematisch-Physikalische Klasse}\ }\textbf {\bibinfo {volume} {1897}},\
  \bibinfo {pages} {71} (\bibinfo {year} {1897})}\BibitemShut {NoStop}%
\bibitem [{\citenamefont {Reck}\ \emph {et~al.}(1994)\citenamefont {Reck},
  \citenamefont {Zeilinger}, \citenamefont {Bernstein},\ and\ \citenamefont
  {Bertani}}]{Reck1994}%
  \BibitemOpen
  \bibfield  {author} {\bibinfo {author} {\bibfnamefont {M.}~\bibnamefont
  {Reck}}, \bibinfo {author} {\bibfnamefont {A.}~\bibnamefont {Zeilinger}},
  \bibinfo {author} {\bibfnamefont {H.~J.}\ \bibnamefont {Bernstein}}, \ and\
  \bibinfo {author} {\bibfnamefont {P.}~\bibnamefont {Bertani}},\ }\href
  {\doibase 10.1103/PhysRevLett.73.58} {\bibfield  {journal} {\bibinfo
  {journal} {Physical Review Letters}\ }\textbf {\bibinfo {volume} {73}},\
  \bibinfo {pages} {58} (\bibinfo {year} {1994})}\BibitemShut {NoStop}%
\end{thebibliography}%
\clearpage 


\end{document}